\documentclass[12pt]{article}
\usepackage{amssymb}
\usepackage{amsmath}
\usepackage{cite}
\usepackage{tensor}
\usepackage{hyperref}
\usepackage{amsfonts}
\usepackage{soul}
\usepackage{comment}
\usepackage{cleveref}
\usepackage{breqn}
\usepackage{mathtools}
\usepackage{tikz-cd}
\usepackage{bbm}

\newcommand{\cat}{\mathcal{C}}
\newcommand{\C}{\mathbb{C}}

\newcommand{\Z}{\mathbb{Z}}
\input xy
\xyoption{all}
\usepackage{enumitem}

\def\sqr#1#2{{\vcenter{\vbox{\hrule height.#2pt
            \hbox{\vrule width.#2pt height#1pt \kern#1pt
                  \vrule width.#2pt}\hrule height.#2pt}}}}

\def\sqra#1#2#3{{\vcenter{\vbox{\hrule height.#2pt
            \hbox{\vrule width.#2pt height#1pt \kern5pt %\kern#1pt
#3
                  \vrule width.#2pt}\hrule height.#2pt}}}}

\numberwithin{equation}{section}

\numberwithin{table}{section}

\begin{document}

\begin{center}

{\large\bf Discrete torsion in gauging non-invertible symmetries}

\vspace*{0.2in}

Alonso Perez-Lona

\begin{tabular}{l}
Department of Physics, MC 0435\\
850 West Campus Drive\\
Virginia Tech\\
Blacksburg, VA  24061
\end{tabular}

{\tt aperezl@vt.edu}
$\,$

\end{center}

In this paper we discuss generalizations of discrete torsion to noninvertible symmetries in 2d QFTs.  One point of this paper is to explain that there are two complementary generalizations. Both generalizations are counted by $H^2(G,U(1))$ when one specializes to ordinary finite groups $G$. However, the counting is different for more general fusion categories. Furthermore, only one generalizes the picture of discrete torsion as differences in choices of gauge actions on B fields. Explaining this in detail, how one of the generalizations of discrete torsion to noninvertible cases encodes actions on B fields, is the other point of this paper. In particular, this generalizes old results in ordinary orbifolds that discrete torsion is a choice of group action on the B field. We also explain how this same generalization of discrete torsion gives rise to physically-sensible twists on gaugeable algebras and fiber functors.

\begin{flushleft}
July 2024
\end{flushleft}

\newpage

\tableofcontents

\newpage

\section{Introduction}

The understanding of what is known as ``discrete torsion''  has significantly evolved since its discovery in \cite{Vafa:1986wx}. In that article, discrete torsion was originally described as an ambiguity in (consistent) phases of orbifold partition functions, where different choices yield physically distinct theories. Later, in \cite{Sharpe:1999pv,Sharpe:1999xw,Sharpe:2000ki}, these phases were identified as (differences of) choices of actions of the orbifold group on B fields. In both cases, discrete torsion has a straightforward classification as cohomology classes in the second cohomology group $H^2(G,U(1))$. We also explain their effect in partition functions. More recently \cite{Brunner:2014lua}, and particularly in the context of generalized symmetries  and their gauging via Frobenius algebras, discrete torsion has been characterized (see \cite[Equation 4.4]{Bhardwaj:2017xup}, \cite[Section 4.2]{Putrov24}) as part of a choice of a Frobenius algebra. In more detail, given an object $A$ in the symmetry category $\cat$, one regards the different possible choices of (special symmetric Frobenius) algebra structures $(\mu,\Delta)$ on a single $A$ as encoding a choice of discrete torsion on that $A$. Specializing to the regular object in $\cat$, discrete torsion has equivalently been characterized as a choice of a fiber functor (see e.g. \cite[slide 26]{Thornslides}). These descriptions are more widely applicable but, curiously, in general only form a set.

One goal of this article is to highlight how generalizing to non-invertible symmetries actually gives rise to two different but complementary notions of discrete torsion: discrete torsion \textit{choices}, and discrete torsion \textit{twists}. We explicitly describe how the \textit{set} of \textit{choices} admits a consistent \textit{twisting action} of the \textit{cohomology group} of \textit{twists} classified by an appropriate notion of cohomology of fusion categories. This is summarized as follows:
\begin{eqnarray*}
    \text{twisting action}: \{\text{d.t.}\,\textit{twists}\} \times \{\text{d.t.}\,\textit{choices} \}\xrightarrow[]{} \{\text{d.t.}\,\textit{choices} \}.
\end{eqnarray*}
In Section \ref{sec:alternatives}, we define these two generalizations. For ordinary groups, both are counted by $H^2(G,U(1))$, but for more
general non-invertible symmetries, the number of discrete torsion twists is different
from the number of discrete torsion choices. We explain in Section~\ref{sec:intrinsic} how discrete torsion twists on gaugeable algebras are naturally captured by the \textit{lazy cohomology group} $H^2_{\ell}(\mathcal{C})$ of a monoidal category $\mathcal{C}$, as introduced in \cite{PSV10}, where, in the present case, $\mathcal{C}$ plays the role of the acting symmetry category. In particular, when the monoidal category $\mathcal{C}$ is of the form $\text{Rep}(\mathcal{H}^*)$ for $\mathcal{H}$ some (finite-dimensional semi-simple) Hopf algebra, this reduces to the more standard notion of lazy cohomology $H^2_{\ell}(\mathcal{H})$ of a Hopf algebra \cite{BC06}. In turn, when $\mathcal{H}=\C[G]$ is the group algebra of some finite group $G$, this further reduces to the second group cohomology group $H^2_{\rm grp}(G,U(1))$. 

The other primary goal of this article is to explain how the discrete torsion twists (classified by lazy cohomology) can be understood as (differences of) noninvertible actions on B fields, which we explain in Section~\ref{sec:gerbes}.  This connects to and generalizes  the previous characterizations \cite{Sharpe:1999pv,Sharpe:1999xw,Sharpe:2000ki} of discrete torsion in terms of equivariant structures (gauge actions) on line bundle gerbes with connection. This is analogously summarized as follows:
\begin{eqnarray*}
    \text{twisting action}: \{\text{d.t.}\,\textit{twists}\} \times \{\text{action functors}\}\xrightarrow[]{} \{\text{action functors}\}.
\end{eqnarray*}

An important distinction is that while discrete torsion choices only exist in non-anomalous cases, discrete torsion twists are always well-defined. These twists remain meaningful since they now act on anomalous actions and functors, as we describe in particular in Section~\ref{ssec:anomalies}.

Most of the results in this article that do not make reference to simple objects apply to general monoidal categories. However, the main focus will be on taking $\cat$ to be a fusion category, which is understood as the higher non-invertible generalization of finite groups in the context of generalized symmetries. For this reason, we will not deal with technicalities involving the smoothness of $\cat$ as one would need to do if one wanted to more generally discuss symmetries described by Lie groups and its non-invertible generalizations.

\section{Preliminaries}\label{sec:review}
\subsection{Review of non-invertible gaugings}\label{ssec:noninvgaug}

We first provide a brief overview of gaugings of non-invertible symmetries and their partition functions. We review the role of special symmetric Frobenius algebras, which motivates studying discrete torsion in the context of such algebras. The content of this subsection can be found in \cite{Perez-Lona:2023djo}.

Given a two-dimensional quantum theory $\mathcal{T}$, one may generally speak of its global symmetries. Traditionally, these were taken to be invertible transformations and, if some of them are parameterized by a group $G$, one refers to them as $G$-symmetries. More recently, this notion has been generalized \cite{Gaiotto:2014kfa,Sharpe:2015mja} in two major ways: as non-invertible transformations, and as higher transformations, in the sense of higher category theory. Thus, the suitable generalization of $G$-symmetries for $G$ a finite group is symmetries parameterized by a fusion\footnote{This can in turn be generalized to multi-fusion categories, as explored in \cite{JF22}.} category $\cat$, or simply $\cat$-symmetries. The group-like case is recovered by specializing to $\cat=\text{Vec}_G$, the category of $G$-graded complex vector spaces.

Now, in the familiar group-like case, one only talks about gauging a non-anomalous subgroup $H\leq G$ of symmetries. This leads to the notion of a gauge theory, denoted $\mathcal{T}/H$, where one mods out by the $H$-parameterized symmetries. For $\cat$-symmetries, this is generalized to gaugings of subsymmetries, which do not necessarily correspond to fusion subcategories of $\cat$. As it has been discussed in e.g. \cite{Bhardwaj:2017xup,Perez-Lona:2023djo,Diatlyk:2023fwf}, based on the ideas in \cite{Fuchs:2002cm}, these subsymmetries are described by symmetric special Frobenius algebra objects\footnote{To define a symmetric special Frobenius algebra one also needs to specify a unit and a counit. However, we omit these from the notation as it is the multiplication and comultiplication that will play the main role.} $(A,\mu,\Delta)$ in $\cat$. Thus, one may more generally study gauge theories $\mathcal{T}/(A,\mu,\Delta)$ where the action of the subsymmetry $(A,\mu,\Delta)$ has been modded out. If one thinks of the objects in $\cat$ as operators implementing the higher non-invertible transformations in $\mathcal{T}$, then the intuition is that the operator associated with $A$ on the gauge theory $\mathcal{T}/(A,\mu,\Delta)$ is simply the identity operator, so that it acts trivially on $\mathcal{T}/(A,\mu,\Delta)$, as expected. That special symmetric Frobenius algebra objects $(A,\mu,\Delta)$ in $\cat$, or gaugeable algebras for short, describe all the possible subsymmetries of $\cat$ is ultimately the reason why we can study discrete torsion of gaugings by looking at how it interacts with these algebra objects. The specific axioms that define a symmetric special Frobenius algebras can be found in e.g. \cite[Appendix A.2]{Perez-Lona:2023djo}.

A subtlety that will be important later on is that the subsymmetries described by a gaugeable algebra $(A,\mu,\Delta)$ only depend on its Morita equivalence class. This means that if there is another gaugeable algebra $(A',\mu',\Delta')$ whose module category in $\cat$ is equivalent to that of $(A,\mu,\Delta)$, then the corresponding subsymmetries and gaugings are physically-equivalent:
\begin{equation}
    (A,\mu,\Delta)\sim (A',\mu',\Delta') \implies T/(A,\mu,\Delta)\cong T/(A',\mu',\Delta') .
\end{equation}
Thus, while for computations in particular one usually works directly with a gaugeable algebra $(A,\mu,\Delta)$, one should keep in mind that the dependence is only on the Morita equivalence class of $(A,\mu,\Delta)$. In the literature, a gaugeable algebra structure on $A$ (or rather, its Morita equivalence class) is what is called a choice of discrete torsion for $A$ \cite{Diatlyk:2023fwf,Putrov24}.

Relevant to discrete torsion is the partition function of the gauge theory, which we now review. Indeed, discrete torsion for groups was discovered \cite{Vafa:1986wx} as the freedom of multiplying the twisted sectors of the gauge, or orbifold, theory partition function by some $U(1)$ phases. These phases are required to respect some consistency conditions, prominently modular invariance for 2-torus partition functions, and multi-loop factorization. As we shall see later, the same holds for partition functions of non-invertible symmetries.

Given a 2d quantum theory $\mathcal{T}$  with $\mathcal{C}$-symmetry on a closed surface $\Sigma$, the partition function $Z_{\Sigma}$ of the gauge theory $\mathcal{T}/(A,\mu,\Delta)$ for $(A,\mu,\Delta)$ a $\cat$-subsymmetry is computed by first picking a triangulation of $\Sigma$, then associating according to a prescribed way a series of multiplications $\mu$ and comultiplications $\Delta$ of $(A,\mu,\Delta)$ to the dual triangulation of $\Sigma$, and finally expanding the resulting morphism in terms of some chosen basis of Hom-spaces of $\mathcal{C}$. We refer the reader to \cite{Fuchs:2002cm,Bhardwaj:2017xup,Perez-Lona:2023djo} for details. In the simple case where the surface $\Sigma=T^2$ is a 2-torus, and the fusion category is multiplicity-free, meaning that
\begin{equation*}
    \text{dim}(\text{Hom}(L_1\otimes L_2,L_3)) \leq 1,
\end{equation*}
for $L_1,L_2,L_3$ any three simple objects in $\cat$, the partition function can be written as \cite[Section 2.6]{Perez-Lona:2023djo}
\begin{eqnarray}
    Z 
    & = & \sum_{L_1, L_2, L_3} \mu^{L_3}_{L_1,L_2} \, \Delta^{L_2,L_1}_{L_3} \, Z^{L_3}_{L_1,L_2}.
    \label{eq:Z:genl1}
\end{eqnarray}
A similar expression can be written down for fusion categories not satisfying the multiplicity-free condition \cite{PartII}. As explained in \cite[Section 2.8]{Perez-Lona:2023djo}, a consequence of $(A,\mu,\Delta)$ being a symmetric special Frobenius algebra is that the partition function for any surface is independent of the chosen triangulation. In particular, this means that the 2-torus partition function (\ref{eq:Z:genl1}) is modular invariant as long as we only gauge by special Frobenius algebras. This suggests that discrete torsion of non-invertible symmetries may also respect the consistency condition of modular invariance if it either defines or acts on special symmetric Frobenius algebras. We will elaborate on the significance of this distinction in Section~\ref{sec:alternatives}.

\subsection{Group-like case revisited}\label{ssec:gplikecase}

Having covered the basics of gauge theories and their partition functions, we now review discrete torsion of group-like symmetries. The derivation of all the building blocks necessary to gauge this group-like case can be found in \cite[Section 2.11]{Perez-Lona:2023djo}. 

For group-like symmetries, the symmetry category is of the form $\mathcal{C}:=\text{Rep}(\C[G]^*)\cong \text{Vec}_G$, the fusion category of $G$-graded complex vector spaces for $G$ some finite group. Its simple objects are one-dimensional vector spaces $U_g$ labeled by group elements $g\in G$. It is endowed with a monoidal functor
\begin{eqnarray}
    \otimes:& \cat\times\cat\longrightarrow \cat; \label{eq:tensorprodfunctor}
\\
            & U_g\otimes U_h\mapsto U_{gh},
\end{eqnarray}
determined by the multiplication of $G$. To gauge the whole symmetry category, one picks the regular object
\begin{equation}
    R:= \bigoplus_{g\in G} U_g,
\end{equation}
and endows it with a special symmetric Frobenius algebra structure. Choosing basis vectors $g\in G$ for $R$, the Frobenius algebra structure is given as
\begin{eqnarray}
    \mu(g\otimes h) &=& gh \label{eq:gplikemult},\\
    \Delta_F(g) &=& \frac{1}{|G|} \sum_{h\in G} gh\otimes h^{-1}, \label{eq:gplikecomult}\\
    u(1) &=& 1,\\
    u_F(g) &=& |G|\,\delta_{1,g},
\end{eqnarray}
thus determining the multiplication and comultiplication coefficients as
\begin{eqnarray}
    \mu_{g,h}^{gh}=1; & \Delta_{gh}^{g,h}=\frac{1}{|G|},
\end{eqnarray}
where $\mu_{g,h}^{gh}\in \text{Hom}(U_g\otimes U_h,U_{gh})$, and $\Delta_{gh}^{g,h}\in \text{Hom}(U_{gh},U_g\otimes U_h)$.

The partition function for a genus $1$ surface is then computed as
\begin{equation}
    Z_{g=1} = \sideset{}{'}\sum_{g,h} \mu_{g,h}^{gh}\Delta_{gh}^{h,g} \, Z_{g,h} = \frac{1}{|G|} \sideset{}{'} \sum_{g,h} Z_{g,h}
\end{equation}
where the sum is over commuting pairs $(g,h)$. 

Discrete torsion in this familiar case is expressed by multiplying the partial traces $Z_{g,h}$ by some phases $\epsilon_{g,h}$. The consistent phases can be derived by working with another multiplication $\mu_{\omega}:R\otimes R\to R$ defined as
\begin{equation}
    \mu_{\omega}(g\otimes h):= \omega(g,h) gh,\label{eq:gpliketwistedmult}
\end{equation}
where $\omega(g,h)\in Z^2(G,U(1))$ is a 2-cocycle in group cohomology.

The pair $(R,\mu_{\omega})$ is an algebra by virtue of $\omega$ being a normalized 2-cocycle. One then observes that together with the comultiplication $\Delta_{F,\omega}$ defined as
\begin{equation}
    \Delta_{F,\omega}(g):=\frac{1}{|G|}\sum_{h\in G}\frac{1}{\omega(gh,h^{-1})} \, gh\otimes h^{-1}, \label{eq:gpliketwistedcomult}
\end{equation}
the tuple $(R,\mu_{\omega},u,\Delta_{F,\omega},u_F)$ is a symmetric special Frobenius algebra one can gauge by. The genus $1$ partition function that one obtains when gauging by this algebra is
\begin{equation}\label{eq:gptwistedpartftn}
      Z_{g=1} = \sideset{}{'}\sum_{g,h} \mu_{g,h}^{gh}\Delta_{gh}^{h,g} \, Z_{g,h} = \frac{1}{|G|} \sideset{}{'} \sum_{g,h} \frac{\omega(g,h)}{\omega(h,g)}Z_{g,h},
\end{equation}
where we see that the discrete torsion phases are $\epsilon_{g,h}=\frac{\omega(g,h)}{\omega(h,g)}$, recovering their standard form.

\section{The two alternatives: choices versus twists}\label{sec:alternatives}

We now explain the subtlety that ultimately leads to the two complementary generalizations of discrete torsion for non-invertible symmetries presented in this article. The reader will note that there are two different perspectives regarding how one arrives to the new gaugeable algebra (\ref{eq:gpliketwistedmult},~\ref{eq:gpliketwistedcomult}) of group-like symmetries described in Section~\ref{ssec:gplikecase}. One perspective involves first fixing a gaugeable algebra structure $(\mu,\Delta)$ on $R$, and then twisting the algebra and coalgebra structure by a representative $\omega$ of a cohomology class in the cohomology group $H^2(G,U(1))$ to obtain another gaugeable algebra. A different perspective is to simply select some other gaugeable algebra $(\mu',\Delta')$ structure on $R$. Now, in the group-like case, this distinction is not substantial, since all possible choices of algebra structures on $R$ can be obtained via a twisting process on any fixed algebra structure on $R$ by a 2-cocycle. Therefore, in this special case, the collection of gaugeable algebra structures on $R$ actually exhibits a group structure. 

By contrast, in fusion categories of the form $\text{Rep}({\cal H})$ for ${\cal H}$ a (finite-dimensional semisimple) Hopf algebra, the collection of Morita classes of algebra structures on the regular object of $\text{Rep}({\cal H})$ generally does not have a group structure. Indeed, these algebra structures are in one-to-one correspondence with equivalence classes of fiber functors on $\text{Rep}({\cal H})$ \cite{Ostrik} (c.f. \cite[Equation 3.10]{Putrov24}). These, in turn, are known \cite[Proposition 2.12]{yucocycles} to be in one-to-one correspondence with algebra 2-cocycles on ${\cal H}^*$. However, this collection of cocycles is only a group when ${\cal H^*}$ is cocommutative \cite{sweedlercocycles}. As explained there, the issue is mainly that the (convolution) product of such cocycles is not again a cocycle.

That the collection of algebra structures on $R$ is just a set is not the end of the story. This is where the difference of perspectives on discrete torsion mentioned above becomes tangible. The upshot is that the two different generalizations of discrete torsion are:
\begin{itemize}
    \item $\{\text{d.t.}\,\textit{choices} \} $: the Morita equivalence classes of possible symmetric special Frobenius algebra structures $(\mu,\Delta)$ on $R$, which we call the \textit{set} of discrete torsion \textit{choices},
    \item $\{\text{d.t.}\,\textit{twists}\}$: the lazy cohomology $H^2_{\ell}(\cat)$ of the fusion category $\cat$, which we refer to as the \textit{cohomology group} of discrete torsion \textit{twists},
\end{itemize}
where $H^2_{\ell}(\cat)$, explored in detail in Section~\ref{ssec:lazy2cocycles}, is a cohomology group that is well-defined for any fusion category $\cat$. A crucial point which we also establish in that section (c.f. Equation~(\ref{eq:finaltwistingaction})) is that these two notions, while different, are complementary in the sense that the set of discrete torsion choices admits an action by the cohomology group of discrete torsion twists:
\begin{eqnarray}
    \text{twisting action}:
    &\{\text{d.t.}\,\textit{twists}\} \times \{\text{d.t.}\,\textit{choices} \}\xrightarrow[]{} \{\text{d.t.}\,\textit{choices} \}.\label{eq:dttwistingaction1}
\end{eqnarray}
This twisting action, as we shall see in Sections~\ref{sec:intrinsic} and \ref{sec:gerbes}, respectively, is not exclusive for discrete torsion choices on $R$ but extends to an action on any Morita class of gaugeable algebras in the category $\cat$, and to tensor functors whose domain category is $\cat$.

Let us briefly mention some examples that we will encounter to illustrate how different the collections of discrete torsion twists, and of discrete torsion choices can be.

For group-like symmetries (see Section~\ref{ssec:gplikecase}), the choices of algebra structures on $R$, equivalently fiber functors on $\cat=\text{Vec}_G$, or \textit{discrete torsion choices}, and cohomology twists, or \textit{discrete torsion twists}, are both classified by $H^2(G,U(1))$
\begin{equation}\label{eq:dtgroupmatching}
    \{\text{d.t.}\,\textit{twists}\}\cong \{\text{d.t.}\,\textit{choices}\}\cong H^2(G,U(1)),
\end{equation}
with the twisting action just being the product structure
\begin{eqnarray}
    \times:&  H^2(G,U(1))\times H^2(G,U(1))&\xrightarrow[]{} H^2(G,U(1)).\label{eq:dtgroup}
\end{eqnarray}

A different example is the immediate non-invertible generalization of this setting, fusion categories of the form $\cat=\text{Rep}(H)$ for $H$ a Hopf algebra, for instance the category $\cat=\text{Rep}(\C[D_4])$ (see Sections~\ref{ssec:lazy2cocycles},~\ref{ssec:fiberfunc}). In this case, the set of discrete torsion choices on the regular object $R$ is the Morita equivalence classes of algebra structures on the regular object, equivalently the classes of fiber functors on $\text{Rep}(\C[D_4])$. On the other hand, the group of discrete torsion twists is the lazy cohomology group $H^2_{\ell}(\C[D_4]^*)$ of $\C[D_4]^*$, which is trivial. Clearly, the twisting action as well is trivial.
\begin{eqnarray}
    \{\text{d.t. choices}\} &=& \{ F_1,F_2,F_3\},\\
    \{\text{d.t. twists}\} &=& 1.
\end{eqnarray}
Even before taking equivalence classes on fiber functors, and on lazy 2-cocycles to obtain the set of discrete torsion choices and the group of discrete torsion twists, respectively, we can already see that these two generalizations are different, as follows. The set of fiber functors on $\text{Rep}(\mathcal{H})$ is equivalently \cite[Proposition 2.12]{yucocycles} the set $Z^2(\mathcal{H}^*)$ of Hopf 2-cocycles on $\mathcal{H}^*$. By contrast, lazy 2-cocycles in this case are the group under convolution $(Z^2_{\ell}(\mathcal{H}^*),\star)$ of lazy Hopf 2-cocycles of $\mathcal{H}^*$, whose underlying set $Z^2_{\ell}(\mathcal{H}^*)$ is by definition (e.g. \cite[Equation 1.6]{BK10}) a subset of the Hopf 2-cocycles of $\mathcal{H}^*$.
\begin{eqnarray}
    \{\text{d.t. choices} \}_{\text{rep.}} &=& Z^2(\mathcal{H}^*),
    \\
    \{\text{d.t. twists} \}_{\text{rep.}} &=& Z^2_{\ell}(\mathcal{H}^*),
\end{eqnarray}
where the notation $\{ \ \}_{\text{rep.}}$ is used to denote we have not taken equivalence classes yet. When $\mathcal{H}^*$ is cocommutative, all Hopf 2-cocycles are lazy, so these two characterizations match as in the group-like case.

Yet another example where choices and twists are different are those fusion categories which do not admit a fiber functor, for example the category of representations $\text{Rep}(q\mathcal{H})$ of a quasi-Hopf algebra $q\mathcal{H}$ (see Section~\ref{ssec:anomalies}). In this case, the set of discrete torsion choices on the regular object $R$, i.e. the set of classes of fiber functors, is empty, whereas the group of discrete torsion twists can be non-trivial, as it is defined intrinsically. For example, when $\cat=\text{Vec}^{\alpha}_G$ for $0\neq [\alpha]\in H^3(G,U(1))$ the category of $G$-graded vector spaces with nontrivial associator, there are no fiber functors, yet one can see from the definition of lazy cohomology that the group of discrete torsion twists is simply $H^2(G,U(1))$. Thus
\begin{eqnarray}
    \{\text{d.t. choices}\} &=& \emptyset,
    \\
    \{\text{d.t. twists}\} &=& H^2(G,U(1)).
\end{eqnarray}
Even though there are no discrete torsion choices on the regular object, there is still a twisting action on any of the other gaugeable algebras in $\text{Vec}_G^{\alpha}$ and on the \textit{quasi}-fiber functor that the latter admits.

One simple example\footnote{We thank X. Yu for valuable discussions on this point.} of this situation is the category $\cat=\text{Vec}_{G}^{\alpha}$ with $G=\Z_2^a\times\Z_2^b\times\Z_2^c$ and $\alpha$ the 3-cocycle
\begin{equation}
    \alpha(g_1,g_2,g_3) = \text{exp}(i \pi a_1b_2c_3),
\end{equation}
for $g_i=(a_i,b_i,c_i)\in \Z_2^a\times\Z_2^b\times\Z_2^c$. This 3-cocycle has a nontrivial cohomology class in $H^3(\Z_2^a\times\Z_2^b\times\Z_2^c,U(1))$, and in particular gives rise to a 3d anomaly inflow TFT \cite{Callan:1984sa} (see also e.g. \cite{Yu:2023nyn}). However, the $\Z_2^a\times\Z_2^b$ subgroup is non-anomalous, and admits twists by local counterterms of the form
\begin{equation}
    \omega(g_1,g_2) = \text{exp}(i\pi a_1 b_2),
\end{equation}
which come precisely from the discrete torsion twists in $H^2_{\ell}(\text{Vec}_G^{\alpha})=H^2(\Z_2^a\times\Z_2^b\times\Z_2^c,U(1))=\Z_2^3$.

In the remainder of this paper we will primarily focus on discrete torsion twists.

\section{Discrete torsion twists and partition functions}\label{sec:intrinsic}

In this Section, we describe the precise definition of discrete torsion twists, and its twisting action on gaugeable algebras. In particular, we provide a formula for the twisted partition function. We verify that the same consistency checks originally used for the derivation of discrete torsion of group in \cite{Vafa:1986wx}, modular invariance and multi-loop factorization, also hold here. For the reader's convenience, the definition of 2-cocycles, 2-coboundaries, and cohomology group of fusion categories is collected in Appendix \ref{app:lazycohomology}.

\subsection{Twists in group-like symmetries}

As highlighted in Section~\ref{sec:alternatives}, one way to arrive at the other gaugeable algebra structure (\ref{eq:gpliketwistedmult},~\ref{eq:gpliketwistedcomult}) is by twisting the original algebra structure by a normalized 2-cocycle. We now explain how to understand this process in a way that allows for an immediate generalization to arbitrary fusion categories. The insight is to regard these phases as coming from a \textit{twist} by a natural isomorphism of the tensor product functor $\otimes: \cat\times\cat\to \cat$ (\ref{eq:tensorprodfunctor}). In other words, a natural isomorphism
\begin{equation}\label{eq:dtnatiso}
    \omega: \otimes \Rightarrow \otimes,
\end{equation}
which in components is a collection of natural isomorphisms
\begin{equation}
    \omega_{X,Y}:X\otimes Y\xrightarrow[]{\cong} X\otimes Y,
\end{equation}
for $X,Y\in \text{ob}(\text{Vec}_G)$. We further require that the natural isomorphism $\omega$ satisfies the following two conditions:
\begin{gather}
    \omega_{X,1}=\omega_{1,X} = \text{id}_X; \label{eq:naturalnorm}
    \\
    \omega_{X,Y\otimes Z} \circ (\text{id}_X\otimes \omega_{Y,Z}) = \omega_{X\otimes Y,Z}\circ (\omega_{X,Y}\otimes \text{id}_Z). \label{eq:naturalassoc} 
\end{gather}
 The first condition (\ref{eq:naturalnorm}) is simply a normalization condition, whereas the second condition (\ref{eq:naturalassoc}) states that $\omega$ is associative. Let us briefly comment on the technicalities of the latter condition. This condition involves morphisms on a triple product $X\otimes Y\otimes Z$ and as such should involve the associator $\alpha_{X,Y,Z}:(X\otimes Y)\otimes Z\to X\otimes (Y\otimes Z)$. For simplicity, we will take these associators as understood in all formulae, so that when, for example, we talk about an algebra $A$ having an associative product $\mu:A\otimes A\to A$, we write
\begin{center}
\begin{tikzcd}
A\otimes A\otimes A \arrow[rr, "\text{id}_A\otimes\mu"] \arrow[d, "\mu\otimes\text{id}_A"'] &  & A\otimes A \arrow[d, "\mu"] \\
A\otimes A \arrow[rr, "\mu"']                                                               &  & A                          
\end{tikzcd}
\end{center}
instead of
\begin{center}
\begin{tikzcd}
                                                                                         & A\otimes (A\otimes A) \arrow[rd, "\text{id}_A\otimes\mu"] &                             \\
(A\otimes A)\otimes A \arrow[d, "\mu\otimes\text{id}_A"'] \arrow[ru, "{\alpha_{A,A,A}}"] &                                                           & A\otimes A \arrow[d, "\mu"] \\
A\otimes A \arrow[rr, "\mu"']                                                            &                                                           & A                          
\end{tikzcd}
\end{center}
One should note that the composition $\omega\circ (\omega\otimes\text{id}_{\text{id}_{\mathcal{C}}})$ is an automorphism of the left-product functor $L:=\otimes\circ(\otimes\times\text{id}_{\mathcal{C}})$, whereas the composition $\omega\circ(\text{id}_{\text{id}_{\mathcal{C}}}\times\omega)$ is an automorphism of the right product $R:=\otimes\circ(\text{id}_{\mathcal{C}}\times\otimes)$. The equality of natural transformations described by (\ref{eq:naturalassoc}) should be then understood as
\begin{equation}\label{eq:explicitassoc}
    \alpha\circ (\omega\circ (\omega\otimes\text{id}_{\text{id}_{\mathcal{C}}}))=(\omega\circ(\text{id}_{\text{id}_{\mathcal{C}}}\times\omega))\circ\alpha
\end{equation}
as natural transformations $L\Rightarrow R$, for $\alpha:L\Rightarrow R$ the associator.

In the present case that the monoidal category is $\cat=\text{Vec}_G$ a fusion category, we can restrict our attention to the simple objects $\{U_g\}_{g\in G}$. Moreover, since $\text{Hom}(U_{gh},U_{gh})=\mathbb{C}$, we can identify $\omega_{g,h}$ with a non-zero scalar. Then Equations (\ref{eq:naturalassoc}),(\ref{eq:naturalnorm}) can be written as
\begin{eqnarray}
    \omega_{1,g} = \omega_{g,1} = 1; \label{eq:gpnormalized}
    \\
    \omega_{g,hk}\cdot \omega_{h,k} = \omega_{gh,k}\cdot \omega_{g,h}, \label{eq:gpcocycle}
\end{eqnarray}
which we recognize as the normalized group $2$-cocycle identities, valued in $U(1)\subset \C^{\times}$. This means that the natural transformation $\omega$ is characterized by a normalized group $2$-cocycle $\omega\in Z^2(G,U(1))$. We can now use this functor to twist the multiplication $\mu$ of the regular representation $R$, via precomposition:
\begin{equation}
    \mu_{\omega}:=\mu\circ\omega_{R,R}: R\otimes R\xrightarrow{\omega_{R,R}} R\otimes R\xrightarrow{\mu} R,
\end{equation}
which is the same twisted multiplication defined previously (\ref{eq:gpliketwistedmult}).

Similarly, one can obtain the twisted comultiplication by postcomposing the comultiplication $\Delta_F$ with the inverse natural isomorphism
\begin{equation}
    \Delta_{F,\omega}:= \omega_{R,R}^{-1}\circ \Delta_F: R\xrightarrow{\Delta_F} R\otimes R \xrightarrow{\omega_{R,R}^{-1}} R\otimes R,
\end{equation}
yielding the same twisted comultiplication constructed before (\ref{eq:gpliketwistedcomult}). Then we know that the tuple $(R,\mu_{\omega}=\mu\circ \omega_{R,R},u,\Delta_{F,\omega}=\omega_{R,R}^{-1}\circ \Delta_F,u^o)$ is a special symmetric Frobenius algebra one can gauge by. Note in particular that the (co)unit is not affected, only the (co)multiplication.

\subsection{Twists in non-invertible symmetries}\label{ssec:lazy2cocycles}

Based on the characterization of discrete torsion twists as natural isomorphisms we define these in more general fusion categories\footnote{As mentioned in the Introduction, this definition, as well as most of the following results, apply to general monoidal categories.} $(\mathcal{C},\otimes)$ as natural isomorphisms $\omega:\otimes\Rightarrow\otimes$ satisfying the conditions (\ref{eq:naturalassoc}), (\ref{eq:naturalnorm}), namely
\begin{gather*}
    \omega_{X,1}=\omega_{1,X} = \text{id}_X;
    \\
    \omega_{X,Y\otimes Z} \circ (\text{id}_X\otimes \omega_{Y,Z}) = \omega_{X\otimes Y,Z}\circ (\omega_{X,Y}\otimes \text{id}_Z).
\end{gather*}
Such natural isomorphisms are known as \textit{lazy 2-cocycles} \cite[Definition 2.1]{PSV10}. In this article we will simply refer to them as 2-cocycles.

Before investigating further the properties of the collection of 2-cocycles of a symmetry category, we need to make sure this indeed describes a genuine notion of discrete torsion twists. An obvious requirement is that given any symmetric special Frobenius algebra in the symmetry category, twisting by a 2-cocycle again yields a symmetric special Frobenius algebra one can gauge by. We now establish this result.

Let $\mathcal{C}$ be the symmetry category, $\omega$ a natural transformation as above, and $(A,\mu)$ an algebra object in $\mathcal{C}$. Define a new multiplication $\mu_{\omega}:=\mu\circ\omega_{A,A}$. Naturality of $\omega$ implies the commutative diagrams
\begin{center}
\begin{tikzcd}
A\otimes A\otimes A \arrow[rr, "\text{id}_A\otimes \mu"] \arrow[d, "{\omega_{A,A\otimes A}}"'] &  & A\otimes A \arrow[d, "{\omega_{A,A}}"] &  & A\otimes A\otimes A \arrow[rr, "\mu\otimes \text{id}_A"] \arrow[d, "{\omega_{A\otimes A,A}}"'] &  & A\otimes A \arrow[d, "{\omega_{A,A}}"] \\
A\otimes A\otimes A \arrow[rr, "\text{id}_{A}\otimes\mu"']                                     &  & A\otimes A                             &  & A\otimes A \arrow[rr, "\mu\otimes\text{id}_A"']                                                &  & A\otimes A                            
\end{tikzcd}
\end{center}
so that
\begin{eqnarray*}
    \mu\circ\omega_{A,A} \circ (\text{id}_A\otimes (\mu\circ\omega_{A,A})) &=& \mu \circ (\text{id}_A\otimes\mu)\circ(\omega_{A,A\otimes A}\circ (\text{id}_{A}\otimes \omega_{A,A}))
    \\
    &=& \mu \circ (\mu\otimes \text{id}_A)\circ(\omega_{A,A\otimes A}\circ (\text{id}_{A}\otimes \omega_{A,A}))
    \\
    &=&
    \mu \circ (\mu\otimes \text{id}_A)\circ(\omega_{A\otimes A,A}\circ (\omega_{A,A}\otimes \text{id}_A))
    \\
    &=& \mu\circ\omega_{A,A} \circ ((\mu\circ \omega_{A,A})\otimes \text{id}_A)),
\end{eqnarray*}
where the second equality follows from associativity of $\mu$, and the third equality from the cocycle property (\ref{eq:naturalassoc}). Thus the associativity commutativity diagram holds
\begin{center}
    \begin{tikzcd}
A\otimes A\otimes A \arrow[rr, "{\text{id}_A\otimes (\mu\circ\omega_{A,A})}"] \arrow[d, "{(\mu\circ\omega_{A,A})\otimes \text{id}_A}"'] &  & A\otimes A \arrow[d, "{\mu\circ\omega_{A,A}}"] \\
A\otimes A \arrow[rr, "{\mu\circ\omega_{A,A}}"']                                                                                         &  & A                                             
\end{tikzcd}
\end{center}
As for the unit axiom, we have that
\begin{eqnarray*}
    \mu\circ\omega_{A,A} \circ (u\otimes\text{id}_A) &=& \mu\circ (u\otimes\text{id}_A)\circ \omega_{1,A}
    \\
    &=& \text{id}_A\circ \text{id}_A
    \\
    &=& \text{id}_A,
\end{eqnarray*}
where the first equality follows from naturality of $\omega$, the second by $(A,\mu)$ satisfying the unit axiom and $\omega$ being normalized in the sense of (\ref{eq:naturalnorm}). The right unit axiom follows similarly. This shows that $(A,\mu_{\omega},u)$ is an associative unital algebra object\footnote{In fact, more is true. The natural isomorphism $\omega$ actually defines an endofunctor on the category $\mathcal{C}$-Alg of algebra objects in $\mathcal{C}$. This is \cite[Theorem 4.7]{BB13}.} in $\mathcal{C}$.

Now let $(A,\Delta,u^o)$ be a coalgebra object in $\mathcal{C}$. Define a new comultiplication as $\Delta_{\omega}:=\omega_{A,A}^{-1}\circ \Delta$. From the naturality diagrams
\begin{center}
    \begin{tikzcd}
A\otimes A \arrow[rr, "\text{id}_A\otimes\Delta"] \arrow[d, "{\omega_{A,A}^{-1}}"'] &  & A\otimes A\otimes A \arrow[d, "{\omega_{A,A\otimes A}}^{-1}"] &  & A\otimes A \arrow[rr, "\Delta\otimes\text{id}_A"] \arrow[d, "{\omega_{A,A}}^{-1}"'] &  & A\otimes A\otimes A \arrow[d, "{\omega_{A\otimes A,A}^{-1}}"] \\
A\otimes A \arrow[rr, "\text{id}_A\otimes \Delta"']                            &  & A\otimes A\otimes A                                      &  & A\otimes A \arrow[rr, "\Delta\otimes\text{id}_A"']                             &  & A\otimes A\otimes A                                     
\end{tikzcd}
\end{center}
one deduces that
\begin{eqnarray*}
    (\text{id}_A\otimes (\omega_{A,A}^{-1}\circ \Delta))\circ (\omega_{A,A}^{-1}\circ \Delta) &=& (\text{id}_A\otimes\omega_{A,A}^{-1})\circ \omega_{A,A\otimes A}^{-1} \circ (\text{id}_A\otimes \Delta)\circ \Delta 
    \\
    &=& (\text{id}_A\otimes\omega_{A,A}^{-1})\circ \omega_{A,A\otimes A}^{-1} \circ (\Delta\otimes \text{id}_A)\circ \Delta
    \\
     &=&(\omega_{A,A}^{-1}\otimes\text{id}_{A})\circ\omega_{A\otimes A,A}^{-1}\circ (\Delta\otimes\text{id}_A)\circ \Delta
    \\
    &=&
    (\omega_{A,A}^{-1}\otimes\text{id}_A)\circ (\Delta\otimes\text{id}_A) \circ \omega_{A,A}^{-1}\circ\Delta
    \\
    &=&
    ((\omega_{A,A}^{-1}\circ\Delta)\otimes \text{id}_A)\circ (\omega_{A,A}^{-1}\circ\Delta),
\end{eqnarray*}
which establishes coassociativity. Note that we used the fact that (\ref{eq:naturalassoc}) implies that for all $X,Y,Z\in\text{ob}(\mathcal{C})$
\begin{equation}
(\text{id}_X\otimes\omega_{Y,Z}^{-1})\circ\omega_{X,Y\otimes Z}^{-1} = (\omega_{X,Y}^{-1}\otimes\text{id}_Z)\circ \omega_{X\otimes Y,Z}^{-1}.
\end{equation}
Moreover, one has that
\begin{eqnarray*}
    (\text{id}_A\otimes u^o)\circ (\omega_{A,A}^{-1}\circ\Delta) &=& \omega_{A,1}^{-1} \circ (\text{id}_A\otimes u^o)\circ\Delta
    \\
    &=& \text{id}_A\circ\text{id}_A 
    \\ 
    &=& \text{id}_A,
\end{eqnarray*}
thus establishing the (right) counit axiom. This shows that whenever $(A,\Delta,u^o)$ is a counital coassociative coalgebra object in $\mathcal{C}$, so is $(A,\omega_{A,A}^{-1}\circ\Delta,u^o)$.

Now, assume $(A,\mu,u,\Delta,u^o)$ is a symmetric special Frobenius algebra object in $\mathcal{C}$. We first look at the Frobenius identities. One has that
\begin{eqnarray*}
    (\omega^{-1}_{A,A}\circ \Delta) \circ (\mu\circ \omega_{A,A}) &=& \omega^{-1}_{A,A}\circ (\text{id}_A\otimes \mu)\circ (\Delta\otimes\text{id}_A) \circ \omega_{A,A}
    \\
    &=& (\text{id}_A\otimes\mu)\circ\omega^{-1}_{A,A\otimes A} \circ \omega_{A\otimes A,A}\circ (\Delta\otimes\text{id}_A)
    \\
     &=& (\text{id}_A\otimes\mu)\circ (\text{id}_A\otimes\omega_{A,A})\circ(\omega_{A,A}^{-1}\otimes\text{id}_{A})\circ\omega^{-1}_{A\otimes A,A} \circ \omega_{A\otimes A,A}
     \\
     & & 
     \circ (\Delta\otimes\text{id}_A)
     \\
     &=& (\text{id}_A\otimes (\mu\circ\omega_{A,A}) \circ ((\omega_{A,A}^{-1}\circ\Delta)\otimes\text{id}_A).
\end{eqnarray*}
The other Frobenius identity is proven analogously. As for the special conditions, one follows immediately since the unit and counits are not changed, and the other follows as
\begin{eqnarray*}
    (\mu\circ\omega_{A,A})\circ (\omega_{A,A}^{-1}\circ \Delta) &=& \mu \circ (\omega_{A,A}\circ\omega^{-1}_{A,A})\circ \Delta
    \\
    &=&
    \mu\circ\Delta 
    \\
    &=&\text{id}_A.
\end{eqnarray*}
Finally, we look at the symmetric property. Recall \cite[Definition 3.4.ii]{Fuchs:2002cm} that a Frobenius algebra $(A,\mu,u,\Delta,u^o)$ is called symmetric if the following equality of morphisms in $\text{Hom}(A,A^*)$ holds
\begin{equation}
    (\text{id}_{A^*}\otimes (u^o\circ\mu))\circ (\overline{\gamma}_A\otimes\text{id}_A) = ((u^o\circ\mu)\otimes\text{id}_{A^*})\circ(\text{id}_A\otimes\gamma_A),
\end{equation}
for $\overline{\gamma}_A:1\to A^*\otimes A$, $\gamma_A:1\to A\otimes A^*$ the coevaluation maps coming from the pivotal structure of the category $\mathcal{C}$. The $\omega$-twisted symmetric Frobenius algebra $A$ is again a symmetric provided one also twists the coevaluation maps by $\omega$ as
\begin{eqnarray}
    \widetilde{\overline{\gamma}}_A&:=& \omega_{A^*,A}^{-1}\circ \overline{\gamma}_A,\\
    \widetilde{\gamma}_A&:=&\omega_{A,A^*}^{-1}\circ \gamma_A.
\end{eqnarray}
Such a twist must be accompanied by an appropriate twist to the evaluation maps $\epsilon_A:A^*\otimes A\to 1$, $\overline{\epsilon}_A:A\otimes A^*\to 1$ as
\begin{eqnarray}
    \widetilde{\epsilon}_A:=\epsilon_A\circ\omega_{A^*,A},
    \\
    \widetilde{\overline{\epsilon}}_A:= \overline{\epsilon}_A\circ\omega_{A,A^*},
\end{eqnarray}
in order to still have a pivotal structure. More precisely, if $\epsilon,\gamma$ define a pivotal structure, so that they satisfy the identity
\begin{equation}
 (\text{id}_A\otimes \epsilon_A)\circ (\gamma_A\otimes\text{id}_A) = \text{id}_A.
\end{equation}
From the cocycle condition (\ref{eq:naturalassoc}) one sees that
\begin{equation}
   ( \text{id}_A\otimes\omega_{A^*,A})\circ (\omega_{A,A^*}^{-1}\otimes\text{id}_A) = \omega^{-1}_{A,A^*\otimes A}\circ \omega_{A\otimes A^*,A},
\end{equation}
then the $\omega$-twisted maps satisfy
\begin{eqnarray*}
    (\text{id}_A\otimes (\epsilon_A\circ \omega_{A^*,A}))\circ ((\omega_{A,A^*}^{-1}\circ \gamma_A)\otimes\text{id}_A) &=& (\text{id}_A\otimes \epsilon_A)\circ \omega^{-1}_{A,A^*\otimes A}\circ \omega_{A\otimes A^*,A}\circ (\gamma_A\otimes \text{id}_A)
    \\
    &=& \omega^{-1}_{A,1}\circ (\text{id}_A\otimes \epsilon_A)\circ (\gamma_A\otimes \text{id}_A)\circ \omega_{1,A}
    \\
    &=& \text{id}_A,
\end{eqnarray*}
so that we still have a pivotal structure.

Equipped with these $\omega$-twisted coevaluation maps one gets that
\begin{eqnarray*}
    (\text{id}_{A^*}\otimes u^o\circ\mu\circ\omega_{A,A})\circ (\widetilde{\overline{\gamma}}_A\otimes\text{id}_A) &=& (\text{id}_{A^*}\otimes u^o\circ\mu)\circ \omega_{A^*,A\otimes A}^{-1}\circ\omega_{A^*\otimes A,A}\circ (\omega_{A^*,A}\otimes\text{id}_A)
    \\
    & &
    \circ (\widetilde{\overline{\gamma}}_A\otimes\text{id}_A)
    \\
    &=& (\text{id}_{A^*}\otimes u^o\circ\mu)\circ\omega_{A^*\otimes A,A}\circ (\omega_{A^*,A}\otimes\text{id}_A)\circ (\widetilde{\overline{\gamma}}_A\otimes\text{id}_A)
    \\
    &=& (\text{id}_{A^*}\otimes u^o\circ\mu)\circ(\omega_{A^*,A}\otimes\text{id}_A)\circ\omega_{A^*\otimes A,A}\circ  (\widetilde{\overline{\gamma}}_A\otimes\text{id}_A)
    \\
    &=& (\text{id}_{A^*}\otimes u^o\circ\mu)\circ(\omega_{A^*,A}\otimes\text{id}_A)\circ  (\omega_{A^*,A}^{-1}\circ\overline{\gamma}_A\otimes\text{id}_A)
    \\
    &=& (\text{id}_{A^*}\otimes u^o\circ\mu)\circ  (\overline{\gamma}_A\otimes\text{id}_A)
    \\
    &=& ((u^o\circ\mu)\otimes\text{id}_{A^*})\circ(\text{id}_A\otimes\gamma_A)
    \\
    &=& (((u^o\circ\omega_{A,A}\circ\mu)\otimes\text{id}_{A^*})\circ(\text{id}_A\otimes\widetilde{\gamma}_A)).
\end{eqnarray*}
This establishes that given any special Frobenius algebra $(A,\mu,u,\Delta,u^o)$ that is symmetric with respect to coevaluation maps $\gamma$, one can construct another special Frobenius algebra $(A,\mu\circ\omega_{A,A},u,\omega^{-1}_{A,A}\circ\Delta,u^o)$ that is symmetric with respect to coevaluation maps $\widetilde{\gamma}=\omega^{-1}_{A,A^*}\circ\gamma_A$ by twisting the product and coproduct by the cocycle $\omega$:
\begin{equation}\label{eq:explicittwistingaction}
    \text{cocycle twist}: (\omega, (A,\mu,u,\Delta,u^o))\mapsto (A,\mu\circ\omega_{A,A},u,\omega_{A,A}^{-1}\circ\Delta,u^o).
\end{equation}

We highlight that this definition of discrete torsion twists is intrinsic to the symmetry category $\cat$. As we have just shown, a discrete torsion twist acts on any gaugeable algebra $(A,\mu,\Delta)$, meaning an object $A$ together with discrete torsion choice $(\mu,\Delta)$, to obtain another discrete torsion choice on $A$. This, however, does not imply that for a given gaugeable algebra $A$, all of the other possible choices of symmetric special Frobenius structure, or discrete torsion choices, on $A$ come from twisting by the category-theoretic discrete torsion. This is in agreement with the traditional group-like orbifold case. For example, the Klein group $\Z_2\times\Z_2$, being a subgroup of group $Q_8$ of quaternions, describes a gaugeable algebra object $(A,\mu)$ in $\text{Vec}(Q_8)$. Moreover, it admits a different symmetric special algebra structure $(A,\mu')$ since it admits discrete torsion $H^2(\Z_2\times\Z_2,U(1))=\Z_2$. However, a 2-cocycle of the category $\text{Vec}(Q_8)$ is classified by $H^2(Q_8,U(1))=0$, so that no category-theoretic discrete torsion can be used to obtain $\mu'$ from $\mu$. Thus, for categories of the form $\text{Rep}(\mathcal{H}^*)$, these 2-cocycles are more accurately understood as the possible discrete torsion twists for the regular representation in $\text{Rep}(\mathcal{H}^*)$ with its discrete torsion choice given by the fiber functor that reconstructs $\mathcal{H}^*$. This generalizes how $H^2(G,U(1))$ classifies discrete torsion for the regular object $R=\oplus_{g\in G}U_g$ in $\text{Vec}_G$. Nevertheless, even in the Rep(${\cal H})$ case the discrete torsion choices one can endow the regular representation with are generally not all related by the action of discrete torsion twists. We will revisit this phenomenon in more detail in Section \ref{ssec:fiberfunc}.

Having observed that 2-cocycles are a good characterization of discrete torsion twists in the sense that they have a consistent action on discrete torsion choices on any object, we proceed to describe the structure of the collection of 2-cocycles, which we denote as $Z^2(\mathcal{C})$. The proofs leading to Definition \ref{eq:lazycohomology} are not new and can be found in e.g. \cite{PSV10} but we include them for completeness. We first show that this collection has a group structure. Since $\text{Aut}(\otimes)$ is a group, we just need to show that $Z^2(\mathcal{C})$ is a subgroup:
\begin{itemize}
    \item The identity element is given by $\omega_{X,Y}=\text{id}_{X\otimes Y}$. 
    \item Given $\omega,\eta\in Z^2(\mathcal{C})$, we verify that $(\omega\circ\eta)$ satisfies the cocycle identity (\ref{eq:naturalassoc}). This is shown as
    \begin{eqnarray*}
        (\omega\circ\eta)_{X,Y\otimes Z}\circ (\text{id}_X\otimes (\omega\circ\eta)_{Y,Z}) &=& \omega_{X,Y\otimes Z} \circ \eta_{X,Y\otimes Z} \circ (\text{id}_{X}\otimes \omega_{Y,Z})\circ (\text{id}_X\otimes\eta_{Y,Z})
        \\
        &=& \omega_{X,Y\otimes Z}  \circ (\text{id}_{X}\otimes \omega_{Y,Z}) \circ \eta_{X,Y\otimes Z}\circ (\text{id}_X\otimes\eta_{Y,Z})
        \\
        &=& \omega_{X\otimes Y,Z}\circ (\omega_{X,Y}\otimes\text{id}_Z)\circ \eta_{X\otimes Y,Z}\circ (\eta_{X,Y}\otimes\text{id}_Z)
        \\
        &=& (\omega\circ\eta)_{X\otimes Y,Z}\circ ((\omega\circ\eta)_{X,Y}\otimes\text{id}_Z).
    \end{eqnarray*}
    \item The inverse $\omega^{-1}$ of $\omega\in Z^2(\mathcal{C})$ is defined component-wise as $(\omega^{-1})_{X,Y}:=\omega^{-1}_{X,Y}$. Taking the inverse of the cocycle identity satisfied by $\omega$ gives
    \begin{eqnarray*}
(\text{id}_X\otimes\omega_{Y,Z}^{-1})\circ\omega_{X,Y\otimes Z}^{-1} &=& (\omega_{X,Y}^{-1}\otimes\text{id}_Z)\circ \omega_{X\otimes Y,Z}^{-1}
        \\
        \omega^{-1}_{X,Y\otimes Z}\circ (\text{id}_X\otimes\omega_{Y,Z}^{-1}) &= &\omega^{-1}_{X\otimes Y}\circ (\omega_{X,Y}^{-1}\otimes\text{id}_Z),
    \end{eqnarray*}
    where the second step follows from $\omega^{-1}$ being a natural transformation.
\end{itemize}
This shows that discrete torsion twists $Z^2(\mathcal{C})$ form a group, much like in the group-like orbifold case. 

Now, one important property of discrete torsion of group orbifolds is that physically-equivalent choices are not directly classified by cocycles but by cohomology classes in group cohomology $H^2(G,U(1))$, where group 2-cocycles differing by a group 2-coboundary are identified. The generalization of a 2-coboundary in the present context is defined as follows, whose collection turns out to be not only a normal but a central subgroup of $Z^2(\mathcal{C})$ \cite[Prop. 2.7.iii]{PSV10}. As we show momentarily, these coboundaries also represent physically-indistinguishable twists. 

A \textit{2-coboundary} $d\phi$ is a 2-cocycle constructed component-wise as
\begin{equation}
    (d\phi)_{X,Y}:=\phi^{-1}_{X\otimes Y}\circ (\phi_X\otimes\phi_Y),
\end{equation}
where $\phi\in\text{Aut}(\text{id}_{\mathcal{C}})$ is an automorphism
\begin{equation}
    \phi: \text{id}_{\mathcal{C}}\to \text{id}_{\mathcal{C}}.
\end{equation}
First of all, we check that $d\phi$ is a natural isomorphism. For $f:X\to X'$ and $g:Y\to Y'$ we have by naturality of $\phi$:
\begin{eqnarray*}
    (d\phi)_{X',Y'}\circ (f\otimes g) &=& \phi^{-1}_{X'\otimes Y'}\circ(\phi_{X'}\otimes\phi_{Y'})\circ(f\otimes g)
    \\
    &=& \phi^{-1}_{X'\otimes Y'}\circ(f\otimes g)\circ(\phi_{X}\otimes\phi_{Y})
    \\
    &=& (f\otimes g)\circ \phi^{-1}_{X\otimes Y}\circ\circ(\phi_{X}\otimes\phi_{Y})
    \\
    &=& (f\otimes g)\circ (d\phi)_{X,Y}.
\end{eqnarray*}
One can also show that $d\phi$ is a normalized 2-cocycle. The normalization is easily checked as
\begin{eqnarray*}
        d\phi_{X,1} &=& \phi^{-1}_{X\otimes 1}\circ (\phi_X\otimes \phi_1)
        \\
        &=& \phi^{-1}_X\circ\phi_X = \text{id}_X,
\end{eqnarray*}
while the 2-cocycle condition follows as
\begin{eqnarray*}
    d\phi_{X,Y\otimes Z}\circ (\text{id}_X\otimes d\phi_{Y,Z}) &=& \phi^{-1}_{X\otimes (Y\otimes Z)}\circ (\phi_X\otimes\phi_{Y\otimes Z})\circ (\text{id}_X\otimes\phi^{-1}_{Y\otimes Z}\circ (\phi_Y\otimes\phi_Z))
    \\
    &=& \phi^{-1}_{X\otimes Y\otimes Z}\circ (\phi_X\otimes\phi_Y\otimes\phi_Z)
    \\
    &=& \phi^{-1}_{X\otimes Y\otimes Z}\circ (\phi_X\otimes\phi_Y\otimes\phi_Z)
    \\
    &=& \phi_{X\otimes Y\otimes Z}^{-1} \circ (\phi_{X\otimes Y}\otimes\phi_Z)\circ (\phi^{-1}_{X\otimes Y}\circ (\phi_X\otimes \phi_Y)\otimes\text{id}_{Z})
    \\
    &=& d\phi_{X\otimes Y,Z}\circ (d\phi_{X,Y}\otimes\text{id}_Z).
\end{eqnarray*}
Finally, we show that $B^2(\mathcal{C})$ is the center of $Z^2(\mathcal{C})$. For $\omega\in Z^2(\mathcal{C})$ and $d\phi\in B^2(\mathcal{C})$, one observes that, by naturality of $\phi$ and $\omega$, respectively,
\begin{eqnarray*}
    (\omega\circ d\phi)_{X,Y} &=& \omega_{X,Y}\circ \phi^{-1}_{X\otimes Y}\circ (\phi_X\otimes\phi_Y)
    \\
    &=& \phi^{-1}_{X\otimes Y}\circ \omega_{X,Y}\circ(\phi_X\otimes\phi_Y)
    \\
    &=& \phi^{-1}_{X\otimes Y}\circ(\phi_X\otimes\phi_Y)\circ \omega_{X,Y}\\
    &=& (d\phi\circ\omega)_{X,Y}.
\end{eqnarray*}
Thus $B^2(\mathcal{C})$ is a central subgroup of $Z^2(\mathcal{C})$, and thus the quotient
\begin{equation}\label{eq:lazycohomology}
    H^2_{\ell}(\mathcal{C}) := Z^2(\mathcal{C})/B^2(\mathcal{C}),
\end{equation}
is well-defined. The group $H^2_{\ell}(\mathcal{C})$ is known as the \textit{lazy cohomology group} of $\mathcal{C}$ \cite{PSV10}.

Parenthetically, we note that these concepts can also be understood in terms of monoidal functors, which we review in Section \ref{ssec:monoidalfunc}. Indeed, a 2-cocycle $\omega$ is equivalently a monoidal functor
\begin{equation}\label{eq:cocycleasfunctor}
(F=\text{Id}_{\cat},J):\cat\to\cat,
\end{equation}
where the underlying functor (\ref{eq:monoidalfunctor}) is $F=\text{Id}_{\cat}$, and the monoidal structure (\ref{eq:jnaturaliso}) is $J=\omega$. Moreover, a 2-coboundary is a monoidal autoequivalence $(\text{Id}_{\cat},\omega)$ equipped with a monoidal natural isomorphism $\phi:(\text{Id}_{\cat},I)\to (\text{Id}_{\cat},\omega)$, where $I_{X,Y}=\text{id}_{X\otimes Y}$. Finally, the lazy cohomology group is the group of equivalence classes of the form $(\text{Id}_{\cat},\omega)$ modulo monoidal natural isomorphisms. 

We now show that, in direct generalization to the group-like orbifold case, \textit{passing from 2-cocycles to cohomology classes amounts to removing physically-indistinguishable twists}. This follows from the observation that any algebra object $(A,\mu)$ is Morita equivalent to its twist $(A,\mu_{\phi}:=\mu\circ d\phi_{A,A})$ by a 2-coboundary $d\phi_{A,A}$. Then, since partition function computations only depend on the Morita equivalence class of $A$, and not on $A$ itself, we are always guaranteed to obtain physically-equivalent results with and without the $d\phi$-twist.

Showing Morita equivalence between two algebra objects $(A,\mu),(A',\mu')$ in $\cat$ means exhibiting an equivalence of categories of their modules in $\cat$ 
\begin{equation}
    \text{Mod}(\mathcal{C},(A,\mu))\cong \text{Mod}(\mathcal{C},(A',\mu')).
\end{equation}
We now show that an equivalence between $\text{Mod}(\mathcal{C},(A,\mu))$ and $\text{Mod}(\mathcal{C},(A,\mu_{\phi}))$ is realized by the functor
\begin{eqnarray}
    F:&\text{Mod}(\mathcal{C},(A,\mu)) \to \text{Mod}(\mathcal{C},(A,\mu_{\phi}))
    \\
    :& ((M,\rho)\xrightarrow{f} (N,\sigma))\mapsto ((M,\rho_{\phi})\xrightarrow{f} (N,\sigma_{\phi}))
\end{eqnarray}
for $\rho_{\phi}:= \rho\circ (\phi_A\otimes\text{id}_M)$ for any $(A,\mu)$-module $(R,\rho)$.

Suppose $M$ is a left $(A,\mu)$-module $\rho:A\otimes M\to M$. This means that the following equation holds
\begin{equation}
    \rho\circ (\mu\otimes\text{id}_M) = \rho\circ (\text{id}_A\otimes\rho).
\end{equation}

Define a new morphism $\rho_{\phi}:=\rho\circ (\phi_A\otimes \text{id}_M):A\otimes M\to M$. Then
\begin{eqnarray*}
    \rho_{\phi}\circ (\mu_{\phi}\otimes\text{id}_M) &=& \rho\circ (\phi_A\otimes \text{id}_M)\circ (\mu\circ\phi_{A\otimes A}^{-1}\circ (\phi_A\otimes \phi_A) \otimes \text{id}_M)
    \\
    &=& \rho\circ  (\mu\circ (\phi_A\otimes \phi_A) \otimes \text{id}_M)
    \\
    &=& \rho\circ  (\mu \otimes \text{id}_M)\circ ((\phi_A\otimes \phi_A)\otimes\text{id}_M)
    \\
     &=& \rho\circ (\text{id}_A\otimes\rho)\circ ((\phi_A\otimes \phi_A)\otimes\text{id}_M)
     \\
     &=& \rho\circ(\phi_A\otimes\text{id}_M)\circ(\text{id}_A\otimes\rho \circ(\phi_A\otimes\text{id}_M))
     \\
     &=&
     \rho_{\phi}\circ(\text{id}_A\otimes\rho_{\phi}),
\end{eqnarray*}
which means that $M$ is equivalently a $(A,\mu_{\phi})$-module with action $\rho_{\phi}$. 

Now, given a morphism $f:(M,\rho)\to (N,\sigma)$ of $(A,\mu)$-modules, that is, a morphism $f:M\to N$ satisfying the identity
\begin{equation}
   \sigma\circ (\text{id}_A\otimes f) = f\circ\rho,
\end{equation}
one observes that
\begin{eqnarray*}
   \sigma_{\phi}\circ (\text{id}_A\otimes f) &=& \sigma\circ (\phi_A\otimes\text{id}_N)\circ (\text{id}_A\otimes f)
   \\
    &=& \sigma\circ (\text{id}_A\otimes f)\circ (\phi_A\otimes\text{id}_M)
    \\
    &=& f\circ\rho\circ (\phi_A\otimes\text{id}_M)
    \\
    &=& f\circ\rho_{\phi},
\end{eqnarray*}
showing that $f:M\to N$ is also a morphism $f:(M,\rho_{\phi})\to (N,\sigma_{\phi})$ of $(A,\mu_{\phi})$-modules.

The functor $F$ is clearly an equivalence of categories, with strict inverse given by
\begin{eqnarray}
    G:&\text{Mod}(\mathcal{C},(A,\mu_{\phi})) \to \text{Mod}(\mathcal{C},(A,\mu))
    \\
    :& ((M,\rho)\xrightarrow{f} (N,\sigma))\mapsto ((M,\rho_{\phi^{-1}})\xrightarrow{f} (N,\sigma_{\phi^{-1}})).
\end{eqnarray}
This establishes that $(A,\mu)$ and $(A,\mu_{\phi})$ are Morita equivalent, thus showing that $H^2_{\ell}(\mathcal{C})$ precisely removes the physically-trivial information of 2-cocycles. 

This means that the cocycle twist action (\ref{eq:explicittwistingaction}) of 2-cocycles descends to a well-defined \textit{twisting action} of the lazy cohomology group $H^2_{\ell}(\cat)$ on Morita equivalence classes of gaugeable algebras, as claimed in (\ref{eq:dttwistingaction1}):
\begin{equation}\label{eq:finaltwistingaction}
    \text{twisting action}: ([\omega],[(A,\mu,u,\Delta,u^o)])\mapsto [(A,\mu\circ\omega_{A,A},u,\omega_{A,A}^{-1}\circ\Delta,u^o)].
\end{equation}

The reader should note that what we just showed is that any gaugeable algebra is always Morita equivalent to its twist by a 2-coboundary, but not that any algebra is Morita inequivalent (hence physically-distinct) to its twist by a 2-cocycle with a nontrivial cohomology class. In particular, the twisting action (\ref{eq:finaltwistingaction}) can have fixed elements. An obvious example is the trivial gaugeable algebra $A=1$, which cannot be twisted nontrivially by any 2-cocycle $\omega$ independently of its cohomology class since $\omega_{1,1}=\text{id}_1$ by the normalization condition (\ref{eq:naturalnorm}). Thus, while twists by 2-coboundaries are always physically indistinguishable, the nontriviality of twists by cocycles with nonzero cohomology class will depend on the particular object it acts on. We will see a similar phenomenon in the context of twists of monoidal functors in Section~\ref{ssec:monoidalfunc}.

To summarize, we first described that there is a consistent definition of 2-cocycles for non-invertible symmetries described by fusion categories. These 2-cocycles always form a group. Then, we showed that these moreover come with a natural action on gaugeable algebra structures, or discrete torsion choices, on any object. Finally, we showed that this action, at the level of Morita equivalence, can only depend on the cohomology class of such 2-cocycles. Thus, we have the claimed action (\ref{eq:dttwistingaction1}) of discrete torsion twists on discrete torsion choices on any object $A$ in the symmetry category:
\begin{eqnarray}
    \text{twisting action} :& \{\cat\text{-d.t.}\, \textit{twists}\}\times \{\text{d.t.}\, \textit{choices}\, \text{on}\, A \}&\xrightarrow{} \{\text{d.t.}\, \textit{choices}\, \text{on}\, A \}
    \\
    & H^2_{\ell}(\cat) \times \{ [A,\mu,\Delta]_{\text{Morita}}\} &\xrightarrow{} \{ [A,\mu,\Delta]_{\text{Morita}}\} \label{eq:moritaaction}
    \\
    & ([\omega], [A,\mu,\Delta]_{\text{Morita}})&\mapsto [A,\mu_{\omega},\Delta_{\omega}]_{\text{Morita}},
\end{eqnarray}
where $\mu_{\omega}:=\mu\circ\omega_{A,A}$ and $\Delta_{\omega}:=\omega_{A,A}^{-1}\circ\Delta$.

The explicit definition of the group $H^2_{\ell}(\mathcal{C})$ for a general monoidal category $\mathcal{C}$ is relatively recent and has not been computed for a wide variety of monoidal categories. However, when $\mathcal{C}$ is of the form $\text{Rep}(\mathcal{H}^*)$ for $\mathcal{H}^*$ (the dual of) a Hopf algebra, the lazy cohomology group of such a representation category coincides with the lazy cohomology group $H^2_{\ell}(\mathcal{H})$ of the Hopf algebra $\mathcal{H}$, which is a more standard notion \cite{BC06} for which some general results are known, see e.g. \cite{GK10}. 

In particular, the lazy cohomology of $\mathcal{C}=\text{Rep}(\C[G]^*)\cong \text{Vec}_G$ is simply the second group cohomology group $H^2(G,U(1))$ with coefficients in $U(1)$, as we previously observed (c.f. Equations \ref{eq:gpnormalized},~\ref{eq:gpcocycle}). Importantly, one should note that, in general,
\begin{equation}
    H^2_{\ell}(\C[G])\neq H^2_{\ell}(\C[G]^*),
\end{equation}
(and more generally $H^2_{\ell}(\mathcal{H})\neq H^2_{\ell}(\mathcal{H}^*)$), which means that discrete torsion twists for $\text{Vec}_G$ is not the same as discrete torsion for $\text{Rep}(G)$. An example of this mismatch is $G=D_4$, where
\begin{eqnarray}
H^2_{\ell}(\text{Vec}_{D_4}) &=& H^2(D_4,U(1)) = \Z_2,\\
H^2_{\ell}(\text{Rep}(D_4)) &=& 0,
\end{eqnarray}
where the cohomology group $H^2_{\ell}(\text{Rep}(D_4))$ is shown to vanish in \cite[Proposition 7.3]{GK10}. We will say more about this example in Section \ref{ssec:fiberfunc}.

Another example is the cohomology group $H^2_{\ell}(\text{Rep}(A_4))$, where $\text{Rep}(A_4)$ is the non-multiplicity-free fusion category of representations of the alternating group of four elements. This cohomology group has been shown \cite[Proposition 7.7]{GK10} to be
\begin{equation}
H^2_{\ell}(\text{Rep}(A_4))=\Z_2,
\end{equation}
which in particular indicates that the fusion category admits two inequivalent fiber functors. This fact will be used in the context of the classification of gaugeable algebras of $\text{Rep}(A_4)$ in \cite{PartII}.

Finally, we highlight that the lazy cohomology group $H^2_{\ell}(\cat)$ of fusion categories is sometimes non-abelian \cite[Proposition 1.4]{GK10}. This is in contrast with group cohomology groups $H^2(G,U(1))$, which are always abelian. As noted in \cite[Section 2]{GK10}, examples of a Hopf algebra whose lazy cohomology is non-abelian are the dual group algebras of the groups $G_{p,m}$ of order $p^{5m}$ for $p$ prime and $m\geq 3$ constructed in \cite{Sah68}. In other words, the fusion categories $\text{Rep}(G_{p,m})$ exhibit discrete torsion twists which form a non-abelian group.

\subsection{Partition functions and multi-loop factorization}

As we saw in the last section, discrete torsion twists are intrinsic to the fusion category $\cat$ and can be identified with its lazy cohomology group $H^2_{\ell}(\cat)$. In the context of non-invertible gaugings, discrete torsion of a class $[\omega]\in H^2_{\ell}(\cat)$ acts via a representative lazy 2-cocycle $\omega\in Z^2(\cat)$ on any given gaugeable algebra $(A,\mu,\Delta)$ by producing another gaugeable algebra $(A,\mu_{\omega},\Delta_{\omega})$ where $\mu_{\omega}=\mu\circ\omega_{A,A}$ and $\Delta_{\omega}=\omega^{-1}_{A,A}\circ\Delta$. This means that when one ``turns on'' discrete torsion, we are more precisely talking about the action of a discrete torsion twists on a fixed gaugeable algebra $(A,\mu,\Delta)$, which is the twisting action of discrete torsion twists on discrete torsion choices on $A$. 

However, in many cases one speaks of turning on discrete torsion for an object $A$ when there is an obvious choice of gaugeable algebra on $A$. This is the case, for example, of group-like symmetries for $A=\oplus_{g\in G} U_g$ the regular object in $\text{Vec}_G$. In that setting, the obvious choice of gaugeable algebra structure $(A,\mu,\Delta)$ is given by (\ref{eq:gplikemult},~\ref{eq:gplikecomult}). Nevertheless, it is equally valid to start with the algebra structure $(A,\mu_{\omega},\Delta_{\omega})$ and turn on discrete torsion, say, by the 2-cocycle $\omega^{-1}$, which clearly returns $(A,\mu,\Delta)$. That is to say, what looks like nontrivial discrete torsion for one choice of algebra may correspond to no discrete torsion for a different choice of algebra. This is clarified by speaking of turning on discrete torsion as a twist action on a discrete torsion choice.

Let us now concentrate on partition functions. Starting with a gaugeable algebra $(A,\mu,\Delta)$ with partition function
\begin{eqnarray*}
    Z 
    & = & \sum_{L_1, L_2, L_3} \mu^{L_3}_{L_1,L_2} \, \Delta^{L_2,L_1}_{L_3} \, Z^{L_3}_{L_1,L_2},
\end{eqnarray*}
a discrete torsion 2-cocycle $\omega$ will change this partition function to
\begin{eqnarray}
    Z^{\omega} 
    & = & \sum_{L_1, L_2, L_3} (\mu\circ \omega_{A,A})^{L_3}_{L_1,L_2} \, (\omega^{-1}_{A,A}\circ\Delta)^{L_2,L_1}_{L_3} \, Z^{L_3}_{L_1,L_2},\label{eq:twistedpartitionftn}
    \\
    & = & \sum_{L_1, L_2, L_3} (\mu_{\omega})^{L_3}_{L_1,L_2} \, (\Delta_{\omega})^{L_2,L_1}_{L_3} \, Z^{L_3}_{L_1,L_2},
\end{eqnarray}
which is computed using the same methods as for any gaugeable algebra. In general, the change of coefficients due to the action of $\omega$ cannot be factored out. For the group-like case, however, this is indeed possible and the expression (\ref{eq:twistedpartitionftn}) returns the familiar twist (\ref{eq:gptwistedpartftn}) by nontrivial phases.

Let us spell out several consequences of this. First, since the partition function is a physical quantity, the formalism of \cite{Fuchs:2002cm} ensures that this too will only depend on the Morita equivalence class of the algebra $(A,\mu_{\omega},\Delta_{\omega})$. In particular, we showed in Section \ref{ssec:lazy2cocycles} that any gaugeable algebra twisted by a 2-coboundary is Morita equivalent to the original algebra, which then means that the partition function too is invariant under twists by 2-coboundaries, as expected.

Second, since the partition function with discrete torsion twist is simply the partition function of another gaugeable algebra $(A,\mu_{\omega},\Delta_{\omega})$, this automatically implies triangulation invariance and in particular modular invariance for the 2-torus, one of the main criteria originally used to constrain the allowed phases.

Finally, we come to multi-loop factorization. This important property of partition functions also played a role in the original derivation of discrete torsion. In theories gauged by group-like symmetries, if we can write a genus $g$ diagram as a product of genus $a, b$ diagrams, factored on the identity, where $a+b = g$, then
multiloop factorization is the statement that the phases assigned to a genus $g$ diagram should match the product of phases corresponding to the two lower-genus diagrams
(up to a factor of $|G|$, which reflects Euler counterterm ambiguities). 

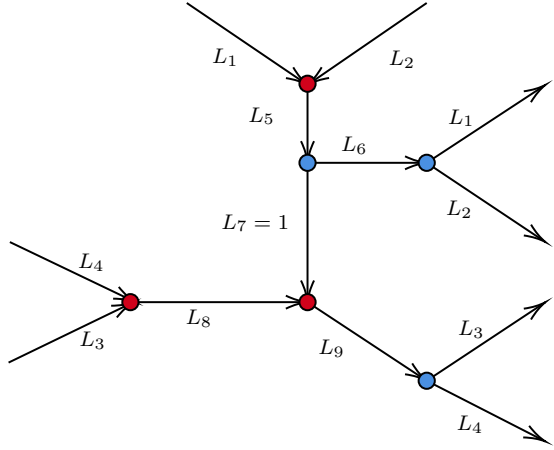
\begin{figure}
    \centering
\tikzset{every picture/.style={line width=0.75pt}} %set default line width to 0.75pt        

\begin{tikzpicture}[x=0.75pt,y=0.75pt,yscale=-1,xscale=1]

%Straight Lines [id:da09879049727609068] 
\draw    (119.44,39) -- (178.68,79.21) ;
\draw [shift={(180.33,80.33)}, rotate = 214.17] [color={rgb, 255:red, 0; green, 0; blue, 0 }  ][line width=0.75]    (10.93,-3.29) .. controls (6.95,-1.4) and (3.31,-0.3) .. (0,0) .. controls (3.31,0.3) and (6.95,1.4) .. (10.93,3.29)   ;
%Straight Lines [id:da022679162896319216] 
\draw    (240.44,39) -- (181.98,79.2) ;
\draw [shift={(180.33,80.33)}, rotate = 325.49] [color={rgb, 255:red, 0; green, 0; blue, 0 }  ][line width=0.75]    (10.93,-3.29) .. controls (6.95,-1.4) and (3.31,-0.3) .. (0,0) .. controls (3.31,0.3) and (6.95,1.4) .. (10.93,3.29)   ;
%Straight Lines [id:da6784132674816423] 
\draw    (180.33,80.33) -- (180.33,117.67) ;
\draw [shift={(180.33,119.67)}, rotate = 270] [color={rgb, 255:red, 0; green, 0; blue, 0 }  ][line width=0.75]    (10.93,-3.29) .. controls (6.95,-1.4) and (3.31,-0.3) .. (0,0) .. controls (3.31,0.3) and (6.95,1.4) .. (10.93,3.29)   ;
%Straight Lines [id:da641402190441402] 
\draw    (180.33,119.67) -- (238.44,119.67) ;
\draw [shift={(240.44,119.67)}, rotate = 180] [color={rgb, 255:red, 0; green, 0; blue, 0 }  ][line width=0.75]    (10.93,-3.29) .. controls (6.95,-1.4) and (3.31,-0.3) .. (0,0) .. controls (3.31,0.3) and (6.95,1.4) .. (10.93,3.29)   ;
%Straight Lines [id:da5715443568775722] 
\draw    (240.44,119.67) -- (298.66,81.43) ;
\draw [shift={(300.33,80.33)}, rotate = 146.7] [color={rgb, 255:red, 0; green, 0; blue, 0 }  ][line width=0.75]    (10.93,-3.29) .. controls (6.95,-1.4) and (3.31,-0.3) .. (0,0) .. controls (3.31,0.3) and (6.95,1.4) .. (10.93,3.29)   ;
%Straight Lines [id:da7770854423418123] 
\draw    (240.44,119.67) -- (298.67,158.56) ;
\draw [shift={(300.33,159.67)}, rotate = 213.74] [color={rgb, 255:red, 0; green, 0; blue, 0 }  ][line width=0.75]    (10.93,-3.29) .. controls (6.95,-1.4) and (3.31,-0.3) .. (0,0) .. controls (3.31,0.3) and (6.95,1.4) .. (10.93,3.29)   ;
%Straight Lines [id:da054430206837292605] 
\draw    (180.33,119.67) -- (180.33,188) ;
\draw [shift={(180.33,190)}, rotate = 270] [color={rgb, 255:red, 0; green, 0; blue, 0 }  ][line width=0.75]    (10.93,-3.29) .. controls (6.95,-1.4) and (3.31,-0.3) .. (0,0) .. controls (3.31,0.3) and (6.95,1.4) .. (10.93,3.29)   ;
%Straight Lines [id:da5048436914641816] 
\draw    (30.22,159.67) -- (91.75,189.14) ;
\draw [shift={(93.56,190)}, rotate = 205.59] [color={rgb, 255:red, 0; green, 0; blue, 0 }  ][line width=0.75]    (10.93,-3.29) .. controls (6.95,-1.4) and (3.31,-0.3) .. (0,0) .. controls (3.31,0.3) and (6.95,1.4) .. (10.93,3.29)   ;
%Straight Lines [id:da7956678657628669] 
\draw    (29.56,220.5) -- (90.76,190.87) ;
\draw [shift={(92.56,190)}, rotate = 154.17] [color={rgb, 255:red, 0; green, 0; blue, 0 }  ][line width=0.75]    (10.93,-3.29) .. controls (6.95,-1.4) and (3.31,-0.3) .. (0,0) .. controls (3.31,0.3) and (6.95,1.4) .. (10.93,3.29)   ;
%Straight Lines [id:da3332475286582053] 
\draw    (92.56,190) -- (178.33,190) ;
\draw [shift={(180.33,190)}, rotate = 180] [color={rgb, 255:red, 0; green, 0; blue, 0 }  ][line width=0.75]    (10.93,-3.29) .. controls (6.95,-1.4) and (3.31,-0.3) .. (0,0) .. controls (3.31,0.3) and (6.95,1.4) .. (10.93,3.29)   ;
%Straight Lines [id:da3810245695382757] 
\draw    (180.33,190) -- (238.78,228.57) ;
\draw [shift={(240.44,229.67)}, rotate = 213.42] [color={rgb, 255:red, 0; green, 0; blue, 0 }  ][line width=0.75]    (10.93,-3.29) .. controls (6.95,-1.4) and (3.31,-0.3) .. (0,0) .. controls (3.31,0.3) and (6.95,1.4) .. (10.93,3.29)   ;
%Straight Lines [id:da8269806107726623] 
\draw    (240.44,229.67) -- (298.67,191.1) ;
\draw [shift={(300.33,190)}, rotate = 146.48] [color={rgb, 255:red, 0; green, 0; blue, 0 }  ][line width=0.75]    (10.93,-3.29) .. controls (6.95,-1.4) and (3.31,-0.3) .. (0,0) .. controls (3.31,0.3) and (6.95,1.4) .. (10.93,3.29)   ;
%Straight Lines [id:da8293040461117034] 
\draw    (240.44,229.67) -- (298.55,259.42) ;
\draw [shift={(300.33,260.33)}, rotate = 207.12] [color={rgb, 255:red, 0; green, 0; blue, 0 }  ][line width=0.75]    (10.93,-3.29) .. controls (6.95,-1.4) and (3.31,-0.3) .. (0,0) .. controls (3.31,0.3) and (6.95,1.4) .. (10.93,3.29)   ;
%Shape: Circle [id:dp7589589714070517] 
\draw  [draw opacity=0][fill={rgb, 255:red, 208; green, 2; blue, 27 }  ,fill opacity=1 ] (176.25,79.75) .. controls (176.25,77.49) and (178.08,75.67) .. (180.33,75.67) .. controls (182.59,75.67) and (184.42,77.49) .. (184.42,79.75) .. controls (184.42,82.01) and (182.59,83.83) .. (180.33,83.83) .. controls (178.08,83.83) and (176.25,82.01) .. (176.25,79.75) -- cycle ;
%Shape: Circle [id:dp6483957670164409] 
\draw  [draw opacity=0][fill={rgb, 255:red, 208; green, 2; blue, 27 }  ,fill opacity=1 ] (86.92,190) .. controls (86.92,187.74) and (88.74,185.92) .. (91,185.92) .. controls (93.26,185.92) and (95.08,187.74) .. (95.08,190) .. controls (95.08,192.26) and (93.26,194.08) .. (91,194.08) .. controls (88.74,194.08) and (86.92,192.26) .. (86.92,190) -- cycle ;
%Shape: Circle [id:dp04093894940371956] 
\draw  [draw opacity=0][fill={rgb, 255:red, 208; green, 2; blue, 27 }  ,fill opacity=1 ] (176.25,190) .. controls (176.25,187.74) and (178.08,185.92) .. (180.33,185.92) .. controls (182.59,185.92) and (184.42,187.74) .. (184.42,190) .. controls (184.42,192.26) and (182.59,194.08) .. (180.33,194.08) .. controls (178.08,194.08) and (176.25,192.26) .. (176.25,190) -- cycle ;
%Shape: Circle [id:dp9448149784544406] 
\draw  [draw opacity=0][fill={rgb, 255:red, 74; green, 144; blue, 226 }  ,fill opacity=1 ] (176.25,119.67) .. controls (176.25,117.41) and (178.08,115.58) .. (180.33,115.58) .. controls (182.59,115.58) and (184.42,117.41) .. (184.42,119.67) .. controls (184.42,121.92) and (182.59,123.75) .. (180.33,123.75) .. controls (178.08,123.75) and (176.25,121.92) .. (176.25,119.67) -- cycle ;
%Shape: Circle [id:dp8577908208543465] 
\draw  [draw opacity=0][fill={rgb, 255:red, 74; green, 144; blue, 226 }  ,fill opacity=1 ] (236.36,119.67) .. controls (236.36,117.41) and (238.19,115.58) .. (240.44,115.58) .. controls (242.7,115.58) and (244.53,117.41) .. (244.53,119.67) .. controls (244.53,121.92) and (242.7,123.75) .. (240.44,123.75) .. controls (238.19,123.75) and (236.36,121.92) .. (236.36,119.67) -- cycle ;
%Shape: Circle [id:dp8330847268561952] 
\draw  [draw opacity=0][fill={rgb, 255:red, 74; green, 144; blue, 226 }  ,fill opacity=1 ] (236.36,229.67) .. controls (236.36,227.41) and (238.19,225.58) .. (240.44,225.58) .. controls (242.7,225.58) and (244.53,227.41) .. (244.53,229.67) .. controls (244.53,231.92) and (242.7,233.75) .. (240.44,233.75) .. controls (238.19,233.75) and (236.36,231.92) .. (236.36,229.67) -- cycle ;

% Text Node
\draw (130.78,60.4) node [anchor=north west][inner sep=0.75pt]  [font=\scriptsize]  {$L_1$};
% Text Node
\draw (219.78,61.4) node [anchor=north west][inner sep=0.75pt]  [font=\scriptsize]  {$L_2$};
% Text Node
\draw (149.11,90.07) node [anchor=north west][inner sep=0.75pt]  [font=\scriptsize]  {$L_5$};
% Text Node
\draw (249.78,91.07) node [anchor=north west][inner sep=0.75pt]  [font=\scriptsize]  {$L_1$};
% Text Node
\draw (248.78,137.4) node [anchor=north west][inner sep=0.75pt]  [font=\scriptsize]  {$L_2$};
% Text Node
\draw (195.78,105.07) node [anchor=north west][inner sep=0.75pt]  [font=\scriptsize]  {$L_6$};
% Text Node
\draw (116.33,127.07) node [anchor=north west][inner sep=0.75pt]  [font=\scriptsize]  {$ $};
% Text Node
\draw (135.44,144.07) node [anchor=north west][inner sep=0.75pt]  [font=\scriptsize]  {$L_7 = 1$};
% Text Node
\draw (116.33,153.07) node [anchor=north west][inner sep=0.75pt]  [font=\scriptsize]  {$ $};
% Text Node
\draw (63.78,203.18) node [anchor=north west][inner sep=0.75pt]  [font=\scriptsize]  {$L_3$};
% Text Node
\draw (63.11,164.07) node [anchor=north west][inner sep=0.75pt]  [font=\scriptsize]  {$L_4$};
% Text Node
\draw (117.44,192.4) node [anchor=north west][inner sep=0.75pt]  [font=\scriptsize]  {$L_8$};
% Text Node
\draw (254.11,245.73) node [anchor=north west][inner sep=0.75pt]  [font=\scriptsize]  {$L_4$};
% Text Node
\draw (254.78,197.73) node [anchor=north west][inner sep=0.75pt]  [font=\scriptsize]  {$L_3$};
% Text Node
\draw (184.33,207.4) node [anchor=north west][inner sep=0.75pt]  [font=\scriptsize]  {$L_9$};

\end{tikzpicture}
    \caption{Example of multiloop factorization at genus 2.  This shows a trivalent resolution of genus 2 surface triangulation in terms of simple lines, which is factored along the line $L_7 = 1$. This diagram defines a genus 2 partial trace, which factors into a product of a pair of genus 1 partial traces defined by the lines above and below $L_7$.}
    \label{fig:g2trivalentgpsimple}
\end{figure}

Here, the same idea applies.  Given a multiloop diagram, expressed as a combination of multiplications $\mu$ and comultiplications $\Delta$, assuming we can restrict the intermediate line to be strictly the identity so that the remaining lines describe a product of lower-genus diagrams, then the phases in the partition function of the whole diagram should be the same as those phases of a product of lower-genus diagrams (up to Euler
counterterm ambiguities). 

Let us analyze how this works for a genus $2$ surface expressed as two 2-tori. The diagram corresponding to the genus $2$ surface in the special case the intermediate line is the identity line is shown in Figure~\ref{fig:g2trivalentgpsimple}. Moreover, as explained in \cite[Section 2.8]{Perez-Lona:2023djo}, its partition function can be read off from it as
\begin{equation}\label{eq:g2partitionfunction}
    Z_{g=2} = \sum_{L_1,\cdots,L_9} \mu_{L_1,L_2}^{L_5}\Delta_{L_5}^{L_6,L_7} \Delta_{L_6}^{L_2,L_1}\mu_{L_3,L_4}^{L_8}\mu_{L_7,L_8}^{L_9}\Delta_{L_9}^{L_4,L_3} Z_{L_1,L_2,L_3,L_4}^{L_5,L_6,L_7,L_8,L_9},
\end{equation}

In the group-like case, the middle line $L_7$, here set to $1$, corresponds to the commutator of the holonomy group elements of the pair of tori. Multiloop factorization in this special case follows as
\begin{eqnarray}
        Z_{g=2} &=& \sum_{a_1,\cdots,b_2} \mu_{a_1,b_1}^{a_1b_1}\Delta_{a_1b_1}^{a_1b_1,1} \Delta_{a_1b_1}^{b_1,a_1}\mu_{a_2,b_2}^{a_2b_2}\mu_{1,a_2b_2}^{a_2b_2}\Delta_{a_2b_2}^{b_2,a_2} Z_{a_1,b_1,a_2,b_2}
        \\
        &=& |G| \sum_{a_1,\cdots,b_2}\left(\mu_{a_1,b_1}^{a_1b_1}\Delta_{a_1b_1}^{a_1b_1,1} \Delta_{a_1b_1}^{b_1,a_1} \right) \left(\mu_{a_2,b_2}^{a_2b_2}\mu_{1,a_2b_2}^{a_2b_2}\Delta_{a_2b_2}^{b_2,a_2}\right) \tilde{Z}_{a_1,b_1,a_2,b_2}
        \\
        &=& \sum_{a_1,\cdots,b_2}\left(\mu_{a_1,b_1}^{a_1b_1}\Delta_{a_1b_1}^{b_1,a_1} \right) \left(\mu_{a_2,b_2}^{a_2b_2}\Delta_{a_2b_2}^{b_2,a_2}\right) \tilde{Z}_{a_1,b_1,a_2,b_2}
\end{eqnarray}
if we take $Z_{a_1,b_1,a_2,b_2}= |G| \,\tilde{Z}_{a_1,b_1,a_2,b_2}$ . This is an Euler counterterm choice, the same choice as in the normalization of orbifold partition functions for genus $g> 1$.

In the more general case, we would still require $L_7=1$, though we still have to sum over the other lines. The partition function becomes
\begin{eqnarray}
    Z_{g=2} &=& \sum_{L_1,\cdots,L_9} \mu_{L_1,L_2}^{L_5}\Delta_{L_5}^{L_6,1} \Delta_{L_6}^{L_2,L_1}\mu_{L_3,L_4}^{L_8}\mu_{1,L_8}^{L_9}\Delta_{L_9}^{L_4,L_3} Z_{L_1,L_2,L_3,L_4}^{L_5,L_6,1,L_8,L_9}\\
    &=&|A| \sum_{L_1,\cdots,L_6,L_8,L_9}\left(\mu_{L_1,L_2}^{L_5}\Delta_{L_5}^{L_6,1} \Delta_{L_6}^{L_2,L_1} \right) \left(\mu_{L_3,L_4}^{L_8}\mu_{1,L_8}^{L_9}\Delta_{L_9}^{L_4,L_3} \right) \tilde{Z}_{L_1,L_2,L_3,L_4}^{L_5,L_6,1,L_8,L_9}
    \\
    &=& \sum_{L_1,\cdots,L_6,L_8,L_9}\left(\mu_{L_1,L_2}^{L_5}\Delta_{L_6}^{L_2,L_1} \right) \left(\mu_{L_3,L_4}^{L_8}\Delta_{L_9}^{L_4,L_3} \right) \tilde{Z}_{L_1,L_2,L_3,L_4}^{L_5,L_6,1,L_8,L_9}
\end{eqnarray}
It is essential for multiloop factorization that $\mu_{1,L}^L=1$ for all
lines $L$. This follows from\footnote{We restrict to the usual case of $\text{dim}(\text{Hom}(1,A))=1$, which justifies the equation $\mu_{1,L}^L=1$.} the unit axiom of the algebra, which forces the coefficient within the algebra to be exactly $1$.  (It is also reflected in sensible choices of intertwiners, which are trivial for the fusion $1 \otimes L \rightarrow L$.)
Similarly, from the counit axiom, $\Delta_L^{L,1}=1/|A|$ for any $L$ (a normalization which fixes the Euler counterterms), and which is also reflected in trivial intertwiners.

All in all, and disregarding the additional factor of $1/|A|$ (e.g.~by defining $Z_{L_1,L_2,L_3,L_4}^{L_5,L_6,L_8,L_9}= |A| \tilde{Z}_{L_1,L_2,L_3,L_4}^{L_5,L_6,L_8,L_9}$) the coefficients multiplying the partial traces become
\begin{equation}
  \epsilon_{L_1,L_2,L_3,L_4}^{L_5,L_6,L_8,L_9} = \epsilon_{L_1,L_2}^{L_5,L_6}\cdot \epsilon_{L_3,L_4}^{L_8,L_9},
\end{equation}
the statement of multi-loop factorization. One should note that this special scenario of $L_7=1$ does not necessarily imply any condition on the fusion coefficients of the $L$'s, it merely requires that coefficients $\Delta_{L_5}^{L_6,L_7}$ or $\mu_{L_7,L_8}^{L_9}$ associated to the Frobenius algebra $\mathcal{A}$ vanish whenever $L_7\neq 1$.

Given that multi-loop factorization is a property of partition functions of gaugeable algebras, Equation (\ref{eq:twistedpartitionftn}) then implies that multi-loop factorization also holds when turning on discrete torsion for any gaugeable algebra.

\section{Discrete torsion twists and B field actions}\label{sec:gerbes}

In Section~\ref{sec:intrinsic}, we saw how the characterization of discrete torsion twists as lazy 2-cocycles gives rise to a natural action on the gaugeable algebras of a given symmetry category. In this section, we show how discrete torsion can be interpreted as the difference between actions on B fields, generalizing old results for orbifolds \cite{Sharpe:1999pv,Sharpe:1999xw,Sharpe:2000ki}. Technically, we describe how the twisting action of 2-cocycles extends to monoidal functors. Since 2-cocycles may be regarded as particular kinds of monoidal autoequivalences, the observation is a special case of the well-known result that a composition of monoidal functors is monoidal again but we provide a proof for completeness. This result is of particular interest when the monoidal functors describe actions on other objects. Indeed, in the most general sense, one can describe the action of a fusion category $\mathcal{C}$ on a mathematical object $\mathcal{V}$ whenever the endomorphisms of $\mathcal{V}$ form a tensor category $\mathcal{E}:=\text{End}_{\otimes}(\mathcal{V})$, by specifying a tensor functor $F:\mathcal{C}\to \mathcal{E}$ (and in fact it is also interesting to consider weakening of this to quasi-tensor functors, as we will also discuss). The object $\mathcal{V}$ may be another fusion category, for example, so that $\mathcal{V}$ becomes a $\mathcal{C}$-module category. But $\mathcal{V}$ may instead be gerbe with connection over some manifold $M$, and the action of $\mathcal{C}$ is described again in terms of a tensor functor.

After presenting the result concerning discrete torssion twists on monoidal functors, we specialize to twists on functors describing the following cases:
\begin{itemize}
    \item fiber functors to $\text{Vec}$ (Section~\ref{ssec:fiberfunc}),
    \item pure gauge non-invertible actions on B-fields (Section~\ref{ssec:actionsgerbes}),
    \item anomalous actions (Section~\ref{ssec:anomalies}).
\end{itemize}

\subsection{Monoidal functors}\label{ssec:monoidalfunc}

Let $\mathcal{C}$ and $\mathcal{E}$ be monoidal categories.  We say that $(F,J)$ is a monoidal functor \cite{EGNO} from $\mathcal{C}$ to $\mathcal{E}$
\begin{equation}\label{eq:monoidalfunctor}
    (F,J): \mathcal{C}\to \mathcal{E},
\end{equation}
if $F$ is a functor of the underlying categories, and $J$ is a natural isomorphism 
\begin{equation}\label{eq:jnaturaliso}
   \begin{tikzcd}
\mathcal{C}\times \mathcal{C} \arrow[rrr, "\otimes_{\mathcal{C}}"] \arrow[dd, "F\times F"']        &  &  & \mathcal{C} \arrow[dd, "F"] \\
                                                                                                   &  &  &                             \\
\mathcal{E}\times \mathcal{E} \arrow[rrr, "\otimes_{\mathcal{E}}"'] \arrow[rrruu, "J", Rightarrow] &  &  & \mathcal{E}                
\end{tikzcd}
\end{equation}
naturally satisfying the following commutative diagram for all objects $X,Y,Z\in\text{ob}(\cat)$:
\begin{equation}   \label{eq:relate-assoc}
\begin{tikzcd}
(F(X)\otimes F(Y))\otimes F(Z) \arrow[rrr, "{a_{F(X),F(Y),F(Z)}}"] \arrow[d, "{J_{X,Y}\otimes \text{id}_{F(Z)}}"'] &  &  & F(X)\otimes (F(Y)\otimes F(Z)) \arrow[d, "{\text{id}_{F(X)}\otimes J_{Y,Z}}"] \\
F(X\otimes Y)\otimes F(Z) \arrow[d, "{J_{X\otimes Y,Z}}"']                                                         &  &  & F(X)\otimes F(Y\otimes Z) \arrow[d, "{J_{X,Y\otimes Z}}"]                     \\
F((X\otimes Y)\otimes Z) \arrow[rrr, "{F(a_{X,Y,Z})}"']                                                            &  &  & F(X\otimes (Y\otimes Z))                                                     
\end{tikzcd}
\end{equation}

Let $(\omega:\otimes\Rightarrow\otimes)\in Z^2(\mathcal{C})$ be a 2-cocycle. We now show that we can define a new monoidal functor $(F',J')$ by setting $F'=F$ and $J'$ computed as the composition of natural isomorphisms
\begin{equation}
\centering
\tikzset{every picture/.style={line width=0.75pt}} %set default line width to 0.75pt        
\begin{tikzpicture}[x=0.75pt,y=0.75pt,yscale=-1,xscale=1]
%uncomment if require: \path (0,211); %set diagram left start at 0, and has height of 211

%Straight Lines [id:da3244508872005072] 
\draw    (236,51.33) -- (236,169.5) ;
\draw [shift={(236,171.5)}, rotate = 270] [color={rgb, 255:red, 0; green, 0; blue, 0 }  ][line width=0.75]    (10.93,-3.29) .. controls (6.95,-1.4) and (3.31,-0.3) .. (0,0) .. controls (3.31,0.3) and (6.95,1.4) .. (10.93,3.29)   ;
%Straight Lines [id:da021582916060037594] 
\draw    (259,183) -- (475.67,183) ;
\draw [shift={(477.67,183)}, rotate = 180] [color={rgb, 255:red, 0; green, 0; blue, 0 }  ][line width=0.75]    (10.93,-3.29) .. controls (6.95,-1.4) and (3.31,-0.3) .. (0,0) .. controls (3.31,0.3) and (6.95,1.4) .. (10.93,3.29)   ;
%Straight Lines [id:da5743447022055379] 
\draw    (259,51.33) -- (475.91,170.54) ;
\draw [shift={(477.67,171.5)}, rotate = 208.79] [color={rgb, 255:red, 0; green, 0; blue, 0 }  ][line width=0.75]    (10.93,-3.29) .. controls (6.95,-1.4) and (3.31,-0.3) .. (0,0) .. controls (3.31,0.3) and (6.95,1.4) .. (10.93,3.29)   ;
%Curve Lines [id:da31106957421450554] 
\draw    (259,33) .. controls (401,40.67) and (401,42) .. (486.67,171.5) ;
\draw [shift={(486.67,171.5)}, rotate = 236.51] [color={rgb, 255:red, 0; green, 0; blue, 0 }  ][line width=0.75]    (10.93,-3.29) .. controls (6.95,-1.4) and (3.31,-0.3) .. (0,0) .. controls (3.31,0.3) and (6.95,1.4) .. (10.93,3.29)   ;
%Right Arrow [id:dp9698540437141896] 
\draw   (346.48,120.42) -- (354.5,116.79) -- (353.01,114.4) -- (361.33,116.76) -- (358.96,123.95) -- (357.48,121.56) -- (349.46,125.19) -- cycle ;
%Shape: Rectangle [id:dp7181003802947907] 
\draw   (249.71,164.23) -- (346.49,120.43) -- (349.46,125.19) -- (252.68,168.99) -- cycle ;
%Shape: Rectangle [id:dp7660057326022995] 
\draw  [color={rgb, 255:red, 255; green, 255; blue, 255 }  ,draw opacity=1 ][fill={rgb, 255:red, 255; green, 255; blue, 255 }  ,fill opacity=1 ] (266.87,157.54) -- (348.05,120.8) -- (349.91,123.78) -- (268.72,160.52) -- cycle ;
%Right Arrow [id:dp9546664194006669] 
\draw   (372.75,65.69) -- (377.69,58.41) -- (375.17,57.15) -- (383.5,54.81) -- (385.24,62.18) -- (382.72,60.92) -- (377.78,68.21) -- cycle ;
%Shape: Rectangle [id:dp9204356321720861] 
\draw   (351.55,96.88) -- (372.75,65.69) -- (377.78,68.21) -- (356.58,99.4) -- cycle ;
%Shape: Rectangle [id:dp47594999136445626] 
\draw  [color={rgb, 255:red, 255; green, 255; blue, 255 }  ,draw opacity=1 ][fill={rgb, 255:red, 255; green, 255; blue, 255 }  ,fill opacity=1 ] (354.25,94.79) -- (375.45,63.59) -- (378.62,65.18) -- (357.43,96.38) -- cycle ;

% Text Node
\draw (215,33.9) node [anchor=north west][inner sep=0.75pt]    {$\mathcal{C} \times \mathcal{C}$};
% Text Node
\draw (216,175.9) node [anchor=north west][inner sep=0.75pt]    {$\mathcal{E} \times \mathcal{E}$};
% Text Node
\draw (480.67,175.9) node [anchor=north west][inner sep=0.75pt]    {$\mathcal{E}$};
% Text Node
\draw (206,107.32) node [anchor=north west][inner sep=0.75pt]  [font=\scriptsize]  {$F\times F$};
% Text Node
\draw (346,187.4) node [anchor=north west][inner sep=0.75pt]  [font=\scriptsize]  {$\otimes _{\mathcal{E}}$};
% Text Node
\draw (320.9,96.82) node [anchor=north west][inner sep=0.75pt]  [font=\scriptsize]  {$F\otimes _{\mathcal{C}}$};
% Text Node
\draw (371.5,34.9) node [anchor=north west][inner sep=0.75pt]  [font=\scriptsize]  {$F\otimes _{\mathcal{C}}$};
% Text Node
\draw (311.35,144.66) node [anchor=north west][inner sep=0.75pt]  [font=\tiny]  {$J$};
% Text Node
\draw (371.69,79.17) node [anchor=north west][inner sep=0.75pt]  [font=\tiny]  {$F\omega $};

\end{tikzpicture}
\end{equation}
with $F\omega$ the whiskering of $\omega$ by $F$ whose components are defined as
\begin{equation}
    (F\omega)_{X,Y} = F(\omega_{X,Y}).
\end{equation}
This means that the components of $J'$ are computed as the composition
\begin{equation}
    J'_{X,Y}: F(X)\otimes_{\mathcal{E}} F(Y) \xrightarrow{J_{X,Y}}F(X\otimes_{\mathcal{C}}Y)\xrightarrow{F(\omega_{X,Y})} F(X\otimes_{\mathcal{C}}Y).
\end{equation}
Defined in this way $J'$ is clearly a natural isomorphism. It remains to show that it is furthermore a monoidal structure, that is, that is satisfies the commutative diagram (\ref{eq:relate-assoc}). This is seen as follows:
\begin{flalign*}
 &J'_{X,Y\otimes Z}\circ (\text{id}_X\otimes J'_{Y,Z})\circ a_{F(X),F(Y),F(Z)}
\\
&=\displaystyle F(\omega_{X,Y\otimes Z})\circ J_{X,Y\otimes Z}\circ (\text{id}_{F(X)}\otimes (F(\omega_{Y,Z})\circ J_{Y,Z}))\circ a_{F(X),F(Y),F(Z)}
    \\
   & = F(\omega_{X,Y\otimes Z})\circ F(\text{id}_X\otimes \omega_{Y,Z})\circ J_{X,Y\otimes Z}\circ (\text{id}_{F(X)}\otimes J_{Y,Z}))\circ a_{F(X),F(Y),F(Z)}
    \\
   & = F(a_{X,Y,Z}\circ\omega_{X\otimes Y,Z}\circ (\omega_{X,Y}\otimes\text{id}_Z)\circ F(a_{X,Y,Z}^{-1}))\circ F(a_{X,Y,Z})\circ J_{X\otimes Y,Z}\circ (J_{X,Y}\otimes \text{id}_{F(Z)})
    \\
   & = F(a_{X,Y,Z})\circ F(\omega_{X\otimes Y,Z})\circ J_{X\otimes Y,Z} \circ ((F(\omega_{X,Y})\circ J_{X,Y})\otimes\text{id}_Z)
    \\
    &= F(a_{X,Y,Z})\circ J'_{X\otimes Y,Z} \circ (J'_{X,Y}\otimes\text{id}_Z).
\end{flalign*}
This follows from using naturality of $J$, as well as the explicit form (\ref{eq:explicitassoc}) of the 2-cocycle identity.
This establishes that for any monoidal functor $(F,J)$ and discrete torsion representative $\omega$, we can define another monoidal functor $(F,J':=F\omega\circ J)$.

Therefore, just as for algebras (\ref{eq:explicittwistingaction}), we have a cocycle twist action on monoidal functors:
\begin{equation}\label{eq:explicittwistingactionfunctors}
    \text{cocycle twist}: (\omega, (F,J))\mapsto (F,F\omega\circ J).
\end{equation}

Now, the twisted monoidal functor explicitly has a different natural transformation whenever $\omega\neq \text{id}_{\otimes}$, yet in the context of gaugeable algebras we saw that 2-coboundaries never carry physically-meaningful information. Here we observe essentially the same result, which says that \textit{a functor and its twist by a 2-coboundary are always monoidally naturally isomorphic. }

The result follows directly from inspecting the definition of monoidal natural isomorphism, which we recall now. Given a pair of monoidal functors
\begin{equation*}
    (F,J), (F',J'): \mathcal{C}\to \mathcal{E},
\end{equation*}
a monoidal natural isomorphism $\eta:(F,J)\to (F',J')$ is \cite{EGNO} a natural isomorphism of the underlying functors $\eta:F\to F'$ such that the following diagram commutes
\begin{equation}\label{eq:monoidnatiso}
    \begin{tikzcd}
F(X)\otimes F(Y) \arrow[rr, "\eta_X\otimes\eta_Y"] \arrow[dd, "{J_{X,Y}}"'] &  & F'(X)\otimes F'(Y) \arrow[dd, "{J'_{X,Y}}"] \\
                                                                            &  &                                             \\
F(X\otimes Y) \arrow[rr, "\eta_{X\otimes Y}"']                              &  & F'(X\otimes Y)                             
\end{tikzcd}.
\end{equation}
Now consider a monoidal functor $(F,J)$ and its twist $(F,Fd\phi\circ J)$ by a 2-coboundary $d\phi$. Then $\eta:=F\phi^{-1}:(F,J)\to (F,Fd\phi\circ J)$ exhibits a monoidal natural isomorphism. It is a natural isomorphism of $F$ since it is the whiskering of a natural isomorphism of $\text{id}_{\mathcal{C}}$. Furthermore, the diagram (\ref{eq:monoidnatiso}) is satisfied as
\begin{eqnarray*}
    (Fd\phi_{X,Y}\circ J_{X,Y})\circ (\eta_X\otimes \eta_Y)&=& Fd\phi_{X,Y}\circ J_{X,Y}\circ (F\phi^{-1}_X\otimes F\phi^{-1}_Y)
    \\
    &=& F\phi^{-1}_{X\otimes Y}\circ F(\phi_X\otimes \phi_Y)\circ J_{X,Y}\circ (F\phi_X\otimes F\phi_Y)
    \\
&=& F\phi^{-1}_{X\otimes Y}\circ J_{X,Y}
\\
&=& \eta_{X\otimes Y}\circ J_{X,Y},
\end{eqnarray*}
by using the naturality of $J$.

This provides the analogue of the twisting action (\ref{eq:finaltwistingaction}) of the lazy cohomology groups but now at the level of monoidal equivalence classes of functors
\begin{eqnarray}
    \text{twisting action} :& ([\omega], [(F,J)])&\mapsto [(F,F\omega\circ J)].
\end{eqnarray}

In many cases, the converse statement, which says that if a functor and its twist are naturally isomorphic then the twisting cocycle is a coboundary,  will hold. For example, in the case of twists of the trivial monoidal autoequivalence $(\text{Id}_{\cat},I):\cat\to\cat$ by a 2-cocycle the converse holds essentially by definition. This is also true for twists of a fiber functor on $\text{Vec}_G$, which we discuss in Section \ref{ssec:fiberfunc}. However, just as for twists of gaugeable algebra, one should be aware that the converse will not hold in general, as one can see from the following example.

Consider the categories $\cat_{\Z_2}^{A}$ whose objects $U_g$ are labeled by the elements $g\in\Z_2$ endowed with a strictly associative monoidal product given by the product on $\Z_2$, and where the hom-spaces are 
\begin{equation}
    \text{Hom}(U_g,U_h)=
    \begin{cases}
      \emptyset, & \text{if}\ g\neq h \\
      A, & g=h
    \end{cases}
  \end{equation}
for $A$ an abelian group. These categories are the ones described in \cite[Example 2.3.6]{EGNO}. Specializing to $A=\Z_2,U(1)$, we have a monoidal functor
\begin{equation}\label{eq:z2functor}
    (F,J):=\cat_{\Z_2}^{\Z_2}\to \cat_{\Z_2}^{U(1)},
\end{equation}
where $F(U_g)=U_g$, and the action of $F$ on the morphisms is given by the injective homomorphism $\imath:\Z_2\hookrightarrow U(1)$. The natural isomorphisms $J_{X,Y}$ are all the identity.

Now, discrete torsion in $\cat_{\Z_2}^{\Z_2}$ is clearly classified by the second group cohomology group $H^2(\Z_2,\Z_2)=\Z_2$. Since the cocycles are normalized, the generator $\omega$ has identity components for all pairs in $\Z_2$ except for
\begin{equation}
    \omega(U_z,U_z)=-\text{id}_{U_z\otimes U_z}.
\end{equation}
Twisting the functor (\ref{eq:z2functor}) by this cocycle changes the monoidal structure $J$ to be the nonidentity morphism for the pair $(U_z,U_z)$ as
\begin{equation}
    J'_{U_z,U_z}=F(\omega_{U_z,U_z})\circ J_{U_z,U_z} = -\text{id}_{U_z\otimes U_z}.
\end{equation}
In group theoretic terms, this can be understood as the image of the 2-cocycle $\omega$ under the pushforward
\begin{equation}
    \imath_*: Z^2(\Z_2,\Z_2)\to Z^2(\Z_2,U(1)),
\end{equation}
induced by the inclusion $\imath$. This pushforward exists because both coefficient groups are taken to be trivial $\Z_2$-modules. The crucial point is that $H^2(\Z_2,U(1))=0$, so that this cocycle must be trivialized. In group cohomology, the image of such cocycle is the coboundary
\begin{equation}
    \imath_*\omega = \delta \phi
\end{equation}
of the 1-chain $\phi:\Z_2\to U(1)$ defined as $\phi(z)=i$. At the level of categories, this trivialization is realized as a monoidal natural isomorphism $\eta:(F,J)\to (F,F(\omega)\circ J)$ whose components are
\begin{eqnarray}
    \eta_{U_1}=\text{id}_{U_1}; & \eta_{U_z}= i\  \text{id}_{U_z},
\end{eqnarray}
since
\begin{eqnarray*}
    J'_{U_z,U_z}\circ (\eta_{U_z}\otimes\eta_{U_z}) &=& F(\omega_{U_z,U_z})\circ J_{U_z,U_z}\circ (\eta_{U_z}\otimes\eta_{U_z}) 
    \\
    &=& \text{id}_{U_z\otimes U_z}\\
    &=& \eta_{U_z\otimes U_z}\circ J_{U_z,U_z}.
\end{eqnarray*}
Thus in this case the original functor is naturally isomorphic to its twist even though the twisting cocycle has a nontrivial cohomology class. This is consistent since the monoidal natural transformation realizing the equivalence between the original functor and the twisted functor does not come from a natural transformation on the source category (e.g. a 2-coboundary), as in particular the complex number $i$ is not part of the morphisms in $\C_{\Z_2}^{\Z_2}$. 

This again shows that while twists by 2-coboundaries are always trivial, in the sense of Morita equivalence for algebras and monoidal natural equivalence for monoidal functors, the nontriviality of a twist by a 2-cocycle with nonzero cohomology class will depend on the specific object it acts on. In particular, there can be monoidal functors that are fixed by the twisting action.

All in all, these observations provide an analogue of the twisting action (\ref{eq:moritaaction}) but now on equivalence classes of monoidal functors:
\begin{eqnarray}
    \text{twisting action} :& \{\cat\text{-d.t.}\, \textit{twists}\}\times \{(F,J):\cat\to\mathcal{E} \}&\xrightarrow{} \{(F,J):\cat\to\mathcal{E} \}
    \\
    & H^2_{\ell}(\cat) \times \{[(F,J):\cat\to\mathcal{E}] \} &\xrightarrow{} \{[(F,J):\cat\to\mathcal{E}] \} \label{eq:functoraction}
    \\
    &(\omega,[(F,J):\cat\to\mathcal{E}]) &\mapsto [(F,F\omega\circ J):\cat\to\mathcal{E}],
\end{eqnarray}
where the equivalence classes are equivalence classes of monoidal naturally isomorphic monoidal functors.

\subsection{Fiber functors}\label{ssec:fiberfunc}
We now apply this result to functors to the category $\mathcal{E}=\text{Vec}$ of vector spaces. This situation in particular describes the relation between discrete torsion twists and fiber functors, which, as discussed in Section~\ref{sec:alternatives}, for the regular object $R$ provide an equivalent characterization of discrete torsion choices on $R$. In particular, a discrete torsion twist generally is not the same as a fiber functor, but acts on these. To make this difference clear, we will also discuss examples with no discrete torsion twists but multiple inequivalent fiber functors, and later in Section~\ref{ssec:anomalies} we will discuss the opposite case, that of discrete torsion twists with no fiber functors.

Assume the fusion category $\cat$ admits an action on $\text{Vec}$. Since $\text{End}_{\otimes}(\text{Vec})\cong\text{Vec}$, a $\cat$-action on $\text{Vec}$ is equivalently a fiber functor $(F,J):\cat\to \text{Vec}$. Then one can twist this fiber functor by a 2-cocycle $\omega$ using the twisting action (\ref{eq:functoraction}) to obtain a new fiber functor $(F,F(\omega)\circ J)$. In particular, by the discussion in Section~\ref{sec:alternatives} this is the specialization of the twisting action (\ref{eq:moritaaction}) on discrete torsion choices on the regular object in $\cat$.

This carries a number of subtleties we think important to highlight. An illustrative example is provided by the category $\text{Rep}(\C[G]^*)=\text{Vec}_G$ of $G$-graded vector spaces with trivial associator. By the Tannaka reconstruction theorem (see e.g. \cite[Theorem 5.2.3]{EGNO}), this category admits at least one fiber functor $(F,J)$, the fiber functor that reconstructs the Hopf algebra as $\text{End}(F,J)=\C[G]^*$. Given that $F(U_g)=\mathbbm{1}\in\text{ob}(\text{Vec})$ for all simples, the natural isomorphisms that characterize the fiber functor are non-zero complex numbers $J_{U_g,U_h}\in\C^{\times}\subset \text{Hom}(\mathbbm{1}\otimes \mathbbm{1},\mathbbm{1})$, or more precisely $U(1)$ phases assuming a normalization condition, required to satisfy the monoidal structure diagram (\ref{eq:relate-assoc}), which is simply the 2-cocycle condition for group 2-cocycles $Z^2(G,U(1))$. Now, in particular, there is the fiber functor $(F,I)$ where one takes $I_{U_g,U_h}:=1$ for all simples. One can see from the defining diagram (\ref{eq:monoidnatiso}) that the 2-cocycle described by any fiber functor $(F,J)$ admitting a monoidal natural isomorphism from $(F,I)$ has a trivial cohomology class, meaning that equivalence classes of monoidal functors are classified by the second group cohomology group $H^2(G,U(1))$. This is, as expected, the same situation encountered at the level of algebras for group-like symmetries (\ref{eq:dtgroupmatching}). However, as we learned, it is more fruitful to keep the distinction between choices of structure (be it algebras, or fiber functors), and twists of these, so that even when the choices no longer form a set, we still have a well-defined twisting action.

A different situation where the distinction between discrete torsion twists and fiber functors is made evident is the case of Tambara-Yamagami (TY) categories \cite{ty}, fusion categories $\cat(A,\chi,\tau)$ whose simple objects are labeled by elements $g$ of an abelian group $A$ and an additional simple $m$, satisfying the fusion rules
\begin{eqnarray*}
    g\otimes h = gh,\\
    g\otimes m = m\otimes g = m,\\
    m\otimes m = \oplus_{g\in A} \ g,
\end{eqnarray*}
and whose associator is determined by a bicharacter $\chi:A\times A\to\C^{\times}$ and a square root $\tau^2=\vert A\vert$. In \cite[Proposition 1]{tambara}, it is shown in particular that any monoidal autoequivalence $(F,J):\cat(A,\chi,\tau)\to \cat(A,\chi,\tau)$ with $F=\text{Id}_{\cat(A,\chi,\tau)}$ is necessarily monoidal naturally isomorphic to the trivial monoidal natural equivalence $(\text{Id}_{\cat(A,\chi,\tau)},I)$, which as discussed in Section~\ref{ssec:lazy2cocycles} (c.f. Equation \ref{eq:cocycleasfunctor}) equivalently means that discrete torsion is always trivial for these categories. However, in \cite[Section 3]{tambara} it is shown that in general these categories admit different equivalence classes\footnote{In this context, two fiber functors are considered equivalent if one can be obtained by precomposing the other with a monoidal autoequivalence of $\cat(A,\chi,\tau)$.} of fiber functors.

The particular case where the fusion category $\cat$ is the TY category (non-canonically) equivalent to $\text{Rep}(D_4)$, the category of representations of the dihedral group of order $8$, connects in a noteworthy way to the discussion on gaugeable objects and discrete torsion in Section~\ref{sec:intrinsic}. The existence of different classes of fiber functors was used in \cite[Section 3.5]{Diatlyk:2023fwf}, based on the observations in \cite[Section 2.4] {Thorngren:2019iar}, to identify the Morita inequivalent gaugeable algebra structures $[\mu,\Delta]$ that the regular object $R$ in $\cat$ admits. The general mechanism that gives rise to this correspondence was described in \cite[Section 2.5]{Perez-Lona:2023djo}: by the Tannaka reconstruction theorem \cite{EGNO}, choosing a fiber functor $(F,J):\cat\to \text{Vec}$ allows to identify $\cat$ with $\text{Rep}(\mathcal{H}_{(F,J)})$, where $\mathcal{H}_{(F,J)}=\text{End}(F,J)$ is the Hopf algebra of endomorphisms of the fiber functor $(F,J)$. Then one can endow the regular object $R$ with a gaugeable algebra structure by constructing a Frobenius algebra from the dual Hopf algebra $\mathcal{H}_{(F,J)}^*$ by following the Larson-Sweedler process \cite{LS69}. Since the vector space of integrals of finite-dimensional Hopf algebras is one-dimensional, this process gives a unique Frobenius algebra for each fiber functor.

Nevertheless, these algebra structures on the regular object of $\text{Rep}(D_4)$, i.e. discrete torsion choices on the regular object, are never obtained one from the other by discrete torsion twists, as all lazy 2-cocycles in this case are 2-coboundaries and thus the twisting action is necessarily trivial. We emphasize that fiber functors and gaugeable algebras are \textit{choices} of structures, whereas a discrete torsion twist is observed through its \textit{action} on such choices. The full relevance of this statement will be observed in Section \ref{ssec:actionsgerbes}, where we properly talk about concrete actions on 2d theories and their relation to discrete torsion. Furthermore, as we have seen, there are cases like that of $\text{Rep}(D_4)$, where there are no discrete torsion twists, but there do exist multiple discrete torsion choices for the regular object, or inequivalent fiber functors. 

Yet a different scenario where this distinction is made explicit is given by relaxing the condition (\ref{eq:relate-assoc}) of monoidal structures. We will return to this in Section \ref{ssec:anomalies}. This series of examples highlights that the classification of discrete torsion choices is generally different from that of discrete torsion twists, in spite of incidental coincidences. Nevertheless, keeping this distinction allows us to preserve a notion of twisting action even when the choices only form a set.

\subsection{Non-invertible gauge actions on B-fields}\label{ssec:actionsgerbes}

We now concentrate on the case where $\mathcal{E}=\text{Mor}(\mathcal{L},\mathcal{L})$ is the category of non-invertible gauge transformations of a B-field on a smooth manifold $M$. Mathematically, this corresponds to the endomorphisms of a line bundle gerbe $\mathcal{L}$ with connection over $M$. In this case, monoidal functors (\ref{eq:monoidalfunctor}), in fact tensor functors, from a fusion category $\mathcal{C}$ to $\mathcal{E}$ describe $\cat$-actions on a gerbe, much as in the case of finite group actions on $U(1)$ bundle gerbes with connection described in \cite{Sharpe:1999pv,Sharpe:1999xw,Sharpe:2000ki}. In particular, this setting shows that the generalization of discrete torsion as lazy cohomology of $\cat$ allows for its explicit application to a wide variety of settings of physical interest.

We emphasize that we regard the B field as being modeled by a \textit{line} bundle gerbe with connection, as opposed to by a \textit{U(1) bundle} gerbe with connection. Though it is true that every gerbe of the latter class gives rise to one of the former class, in direct analogy to how a principal $U(1)$ bundle gives rise to a complex line bundle, we deliberately use this terminology to refer to the fact that we consider the extended notion of morphisms that include non-invertible ones, as first described in \cite{Wald07}, which is a trait of main interest in the context of generalized symmetries. We first review the relevant definitions to make this setting concrete. 

A $U(1)$ bundle gerbe can be thought of as the first higher generalization (in the sense of higher category theory) of a principal $U(1)$ bundle. Even though it is genuinely a higher object, it can be defined in terms of more classical data provided one uses a surjective submersion. For this we follow \cite{Murray:1994db}. Thus, a $U(1)$ bundle gerbe over $M$ consists of:
\begin{itemize}
  \setlength\itemsep{0em}
  \item A surjective submersion $\pi:Y\to M$,
  \item A principal $U(1)$ bundle $P\to Y^{[2]}$ over the fiber product $Y^{[2]}:= \{ (y_1,y_2)\in Y\times Y\vert \pi(y_1)=\pi(y_2)\}$,
  \item A smooth isomorphism of principal $U(1)$ bundles
  \begin{equation}
      \mu_{(y_1,y_2,y_3)}: P_{(y_1,y_2)}\otimes P_{(y_2,y_3)}\xrightarrow{\cong} P_{(y_1,y_3)},
  \end{equation}
  on $Y^{[3]}$ called the \textit{multiplication} morphism, which satisfies the associativity diagram
  \begin{equation}\label{eq:u1gerbeassociativity}
\begin{tikzcd}
{P_{(y_1,y_2)}\otimes P_{(y_2,y_3)} \otimes P_{(y_3,y_4)}} \arrow[rrr, "{\text{id}_{P_{(y_1,y_2)}}\otimes\mu_{(y_2,y_3,y_4)}}"] \arrow[d, "{\mu_{(y_1,y_2,y_3)}\otimes \text{id}_{P_{(y_3,y_4)}}}"'] &  &  & {P_{(y_1,y_2)}\otimes P_{(y_2,y_4)}} \arrow[d, "{\mu_{(y_1,y_2,y_4)}}"] \\
{P_{(y_1,y_3)}\otimes P_{(y_3,y_4)}} \arrow[rrr, "{\mu_{(y_1,y_3,y_4)}}"']                                                                                                                           &  &  & {P_{(y_1,y_4)}}                                                        
\end{tikzcd}
  \end{equation}
\end{itemize}
These are the geometric realizations of the cohomology classes $H^2(M,U(1))$ as follows. Choose a sufficiently fine good cover $\{U_i\}_{i\in I}$. Then there are sections $s_i:U_i\to Y$, and since the cover is good the (pullback) principal $U(1)$ bundle over $\coprod_{ij} U_{ij}$ for $U_{ij}:=U_i\cap U_j$ is trivial. Thus, we can choose sections $\sigma_{ij}:U_{ij}\to P$, such that $\sigma_{ij}(u)\in P_{s_a(u),s_b(u)}$, so that the multiplication morphism $\mu_{ijk}$ satisfies
\begin{equation}
    \sigma_{ij}\sigma_{jk} = \sigma_{ik}\mu_{ijk}.
\end{equation}
Furthermore, the associativity diagram becomes the identity
\begin{equation}
    \mu_{ijl}\mu_{jkl} = \mu_{ikl}\mu_{ijk},
\end{equation}
which means that $\mu_{ijk}$ is a Čech 2-cocycle valued in $U(1)$. The cohomology class $[\mu]$ in $H^3(M,\Z)$ of a gerbe obtained in this way is called its Dixmier-Douady class, which is independent of the choice of good cover and sections. Furthermore, it classifies the gerbe up to (stable) isomorphism, and in fact for any Dixmier-Douady class one can always find a gerbe whose surjective submersion $\pi:Y\to M$ is given by $Y=\coprod_{i\in I}U_i$ an open cover \cite{Murray:1999ew}. Therefore, without loss of generality we can work with gerbes of this form.

We now define a connection on a $U(1)$ bundle gerbe $(Y,P,\mu)$. For this, we define
\begin{equation*}
    Y^{[2]}\circ Y^{[2]}:=\{(x,y)\times (y,z) \in Y^{[2]}\times Y^{[2]}\},
\end{equation*}
and by $P\circ P$ we denote the restriction of the bundle $P\otimes P\to Y^{[2]}\times Y^{[2]}$ to $Y^{[2]}\circ Y^{[2]}$. Let $\nabla$ be a connection on $P$, then this induces a connection $\nabla\circ\nabla$ on $P\circ P$. We say $\nabla$ is a $U(1)$ bundle gerbe connection if its is compatible with the multiplication map $\mu:P\circ P\to P$, that is, the multiplication maps $\nabla\circ\nabla$ to $\nabla$.

At this point, with the appropriate notion of automorphism categories, one can already talk about higher group actions on 2d $\sigma$-models. However, we are interested in also observing non-invertible symmetries. For this reason, we now pass to the setting of line bundle gerbes. In particular for the appropriate notion of morphisms, we follow \cite{Wald07}. An extensive review of this content can also be found in \cite{Bunk:2016rta}.

Recall that any principal $U(1)$ bundle gives rise to a line bundle via its fundamental representation $\rho:U(1)\hookrightarrow {\rm GL}_1(\mathbb{C})=\mathbb{C}^{\times}$. The same is true for $U(1)$ and line bundle gerbes, which explains the similarity in definitions. A line bundle gerbe over a manifold $M$ consists of the following data:
\begin{itemize}
  \setlength\itemsep{0em}
  \item A surjective submersion $\pi:Y\to M$,
  \item A line bundle $L\to Y^{[2]}$ over the fiber product $Y^{[2]}:= \{ (y_1,y_2)\in Y\times Y\, \vert \, \pi(y_1)=\pi(y_2)\}$,
  \item A smooth isomorphism of \textit{line} bundles, the \textit{multiplication} morphism,
  \begin{equation}
      \mu_{(y_1,y_2,y_3)}: L_{(y_1,y_2)}\otimes L_{(y_2,y_3)}\xrightarrow{\cong} L_{(y_1,y_3)},
  \end{equation}
  on $Y^{[3]}$, satisfying the associativity diagram \ref{eq:u1gerbeassociativity} (at the level of line bundle morphisms).
\end{itemize}

As expected, this definition is obtained simply by considering the associated line bundle of the principal $U(1)$ bundle $P\to Y^{[2]}$ that defines a $U(1)$ gerbe over $M$.

Furthermore, the notion of connection on $U(1)$ bundle gerbes extends to line bundle gerbes. A connection $\nabla$ on a line bundle gerbe is defined as a connection on the line bundle $L\to Y^{[2]}$ compatible with the multiplication morphism $\mu$ of line bundles, explicitly
\begin{equation}
    \nabla\circ\mu = \mu\circ (\nabla\otimes 1 + 1\otimes\nabla).
\end{equation}

Since line bundle gerbes are higher objects, they also feature higher morphisms, in the sense that the collection of line bundle gerbes over a space $M$ forms a 2-category (with extra structure). The relevant aspect for the present purpose is the definition of 1- and 2-morphisms between gerbes, which we recall now.

Given a pair of line bundle gerbes with connection $\mathcal{L}_1=(L_1,\mu_1,\nabla_1)$, $\mathcal{L}_2=(L_2,\mu_2,\nabla_2)$, a 1-morphism of line bundle gerbes consists of:
\begin{itemize}
    \item a surjective submersion $\zeta: Z\to Y_1\otimes_M Y_2$,
    \item a hermitian vector bundle $E\to Z$ of rank $n$ with connection,
    \item an isomorphism of hermitian vector bundles with connection over $Z\times_M Z$
    \begin{equation*}
        \alpha: L_{1,(z_1,z_2)}\otimes E_{z_2}\to E_{z_1}\otimes L_{2,(z_1,z_2)},
    \end{equation*}
    satisfying the commutative diagram of bundles over $Z\times_M Z\times_M Z$
    \begin{center}
   \begin{equation}\label{eq:1morassoc}
       \begin{tikzcd}
                                                                                                       & {L_{1,(z_1,z_2)}\otimes L_{1,(z_2,z_3)}\otimes E_{z_3}} \arrow[rd, "{\mu_{1,(z_1,z_2,z_3)}\otimes 1}"] \arrow[ld, "{1\otimes\alpha_{(z_2,z_3)}}"'] &                                                                    \\
{L_{1,(z_1,z_2)}\otimes E_{z_2}\otimes L_{2,(z_2,z_3)}} \arrow[d, "{\alpha_{(z_1,z_2)\otimes1}}"']     &                                                                                                                                                    & {L_{1,(z_1,z_3)}\otimes E_{z_3}} \arrow[d, "{\alpha_{(z_1,z_2)}}"] \\
{E_{z_1}\otimes L_{2,(z_1,z_2)}\otimes L_{2,(z_2,z_3)}} \arrow[rr, "{1\otimes\mu_{2,(z_1,z_2,z_3)}}"'] &                                                                                                                                                    & {E_{z_1}\otimes L_{2,(z_1,z_3)}}                                  
\end{tikzcd}.
   \end{equation}
    \end{center}
\end{itemize}

The composition of 1-morphisms $\mathcal{L}_1\xrightarrow{(E,\alpha)}\mathcal{L}_2\xrightarrow{(E',\alpha')}$ is defined by the surjective submersion $Z\times_{Y_2}Z'\to Y_1\times_M Y_3$, vector bundle $\text{pr}_Z E\otimes \text{pr}_{Z'}E'$, and isomorphism $(\text{pr}_Z^{[2]*}\alpha)\circ (\text{pr}_{Z'}^{[2]*}\alpha')$, for $\text{pr}_{Z^{(')}}:Z\times_{Y_2} Z'\to Z^{(')}$. 

We pause to discuss some points about 1-morphisms before moving on to 2-morphisms. First, note that the 1-morphisms are described by vector bundles of rank $n$ not necessarily $1$. In contrast, $U(1)$ bundle gerbe morphisms are commonly described by principal $U(1)$ bundles (subject to some conditions), which after linearization describe only vector bundles of rank $n=1$. In fact, 1-morphisms of line bundle gerbes are (stable) isomorphisms if and only if their corresponding vector bundles are of rank $1$ \cite[Section 1.3]{Wald07}. However, even though the morphisms with vector bundles of rank $n>1$ are not isomorphisms, they still have \textit{duals}, as described in \cite[Section 1.4]{Wald07}, which in the context of generalized symmetries serves as a suitable weakening of invertibility that partly justifies regarding non-invertible transformations as symmetries. 

A second important fact is that the composition of 1-morphisms as defined here is \textit{strictly} associative \cite[Proposition 1]{Wald07}. We will make use of this fact below when defining fusion category actions.

Moreover, as mentioned in \cite[Example 4.7]{Bunk:2016rta}, a more explicit form of 1-morphisms is given when both line bundle gerbes $(\mathcal{L}_1,\mu_1),(\mathcal{L}_2,\mu_2)$ are subordinated to an open cover $Y_1=Y_2=\coprod_{i\in I}U_i$. In such a case, a 1-morphism consists of vector bundles with connection $E_i\to U_i$ with isomorphisms of vector bundles with connection
\begin{equation}
    \alpha_{ij}:L_{1,ij}\otimes E_j\xrightarrow{\cong} E_i\otimes L_{2,ij}.
\end{equation}
If furthermore the cover is good then these are just transition functions $\alpha_{ij}:U_{ab}\to GL(n)$ (or to $U(n)$ in the case of Hermitian bundles), where the diagram (\ref{eq:1morassoc}) describes twisted vector bundles
\begin{equation}\label{eq:twistedvector}
    \mu_{2,ijk}\alpha_{ij}\alpha_{jk} = \alpha_{ik}\mu_{2,ijk}.
\end{equation}

As is familiar from the literature of Chan-Paton vector bundles on $D$-branes, the notion of twisted vector bundles only makes sense when the difference of Dixmier-Douady classes classified in $H^3(M,\Z)$ of the source and target gerbes is torsional if one wants to keep the rank of the vector bundle finite. In the present article we will only concentrate on the category of morphisms $\mathcal{E}=\text{Mor}(\mathcal{L},\mathcal{L})$, where this issue clearly does not play a role. However, this means that the $\cat$-actions described by tensor functors to $\mathcal{E}$ are pure higher (but potentially non-invertible) gauge transformations of $\mathcal{L}$, implying in particular that $\cat$ acts completely trivially on the base space $M$. A more encompassing notion of action should also involve a nontrivial action on $M$, so that to each object $C$ of the acting fusion category $\mathcal{C}$ one assigns functorially another gerbe $\mathcal{L}_C$ along with a (not necessarily invertible) $1$-morphism of gerbes
\begin{equation}\label{eq:fullaction}
    \eta: \mathcal{L}\to \mathcal{L}_C.
\end{equation}
This last condition ensures that the curvature $H:=\text{curv}(B)\in\Omega^3(M)$ does not change.
For example, in the case of finite group actions, to each group element $g\in G$ the assigned gerbe is simply the pullback gerbe $g^*\mathcal{L}$, where $g$ acts on $M$ via a diffeomorphism. An assignment such as (\ref{eq:fullaction}) may be regarded as the linear non-invertible generalization of the higher group extensions described in \cite{BMS21,BS23}. This generalization would take us too far from the topic of discrete torsion but we hope to return to this in future work.

We now proceed to describe 2-morphisms. Let $Y_{12}=Y_1\times_M Y_2$. We consider triples $(W,\omega,\beta)$ where $\omega:W\to Z_1\times_{Y_{12}} Z_2$ is a surjective submersion, and $\beta: \omega_{1}^* E_1\to \omega_{2}^* E_2$ is a morphism of vector bundles over $W$ satisfying the commutative diagram over $W\times_M W$':
    \begin{equation}\label{eq:2morphisms}
 \begin{tikzcd}
{L_{1,(w_1,w_2)}\otimes E_{1,w_2}} \arrow[rr, "{\alpha_{1,(w_1,w_2)}}"] \arrow[d, "1\otimes \beta_{w_2}"'] &  & {E_{1,w_1}\otimes L_{2,(w_1,w_2)}} \arrow[d, "\beta_{w_1}\otimes 1"] \\
{L_{1,(w_1,w_2)}\otimes E_{2,w_2}} \arrow[rr, "{\alpha_{2,(w_1,w_2)}}"']                                   &  & {E_{2,w_1}\otimes L_{2,(w_1,w_2)}}                                  
\end{tikzcd}       
    \end{equation}
A 2-morphism is an equivalence class of such triples, where any two pairs $(W,\omega,\beta), (W',\omega',\beta')$ are equivalent if there exists a manifold $X$ with surjective submersions to $W$ and $W'$ forming a commutative diagram
\begin{center}
    \begin{tikzcd}
X \arrow[rr] \arrow[dd]   &  & W \arrow[dd, "\omega"] \\
                          &  &                        \\
W' \arrow[rr, "\omega'"'] &  & Z_1\times_{Y_{12}}Z_2 
\end{tikzcd}
\end{center}
such that the pullback to $X$ of $\beta,\beta'$ along such surjective submersions coincides.

One can define horizontal and vertical compositions for 2-morphisms. Not surprisingly, the 1- and 2-morphisms between any two line bundle gerbes form themselves a category. A less trivial fact is that this are furthermore semisimple abelian and cartesian symmetric monoidal categories, where the product corresponds to the direct sum on vector bundles \cite[Theorem 4.27]{Bunk:2016rta}. Given a pair of 1-morphisms $(E_1,\alpha_1,\zeta_1), (E_2,\alpha_2,\zeta_2):\mathcal{L}_i\to\mathcal{L}_j$, their direct sum $(F,\beta,\zeta)$ is given by the direct sum of pullback vector bundles
\begin{equation*}
    F:=\text{pr}_{Z_1}^*E_1\oplus \text{pr}_{Z_2}^*E_2\to Z_{1}\times_{Y_{12}}Z_2=:Z,
\end{equation*}
with isomorphism of vector bundles with connection given by
\begin{equation*}
    \beta_{(z_1,z_2)}:= (d^r_{L_2,(z_1,z_2}))^{-1}\circ (\alpha_{1,(z_1,z_2)}\oplus \alpha_{2,(z_1,z_2)})\circ d^l_{L_1,(z_1,z_2)},
\end{equation*}
for $\alpha_1,\alpha_2$ pulled back to $Z$ from $Z_1,Z_2$, respectively, and the distribution isomorphisms denoted as
\begin{eqnarray*}
    d^l_L:& L\otimes (E_1\oplus E_2)\to (L\otimes E_1)\oplus (L\otimes E_2),
    \\
    d^l_L:& (E_1\oplus E_2)\otimes L \to (E_1\otimes L)\oplus (E_2\otimes L).
\end{eqnarray*}
By unpacking these definitions we can see that the morphism category $\text{Mor}(\mathcal{L},\mathcal{L})$ is simply the category of hermitian vector bundles with connection over $M$:
\begin{equation}
    \mathcal{E}=\text{Mor}(\mathcal{L},\mathcal{L})\cong \text{HVBdl}_{\nabla}(M).
\end{equation}
In particular, the composition of morphisms is given by the tensor product, which is strictly associative \cite[Proposition 1]{Wald07}. Under these two operations the category $\mathcal{E}$ becomes a distributive category.

All in all, to describe the action of a fusion category $\mathcal{C}$ on a 2d $\sigma$-model on $M$, where $\mathcal{C}$ acts as pure higher but not necessarily invertible gauge transformations, trivially on $M$, it suffices to endow $\mathcal{C}$ with a tensor functor 
\begin{equation}
    (F,J):\mathcal{C}\to \mathcal{E}\cong \text{HVBdl}_{\nabla}(M)
\end{equation}
to the category of hermitian vector bundles with connection on $M$. 

We therefore see that, in particular, to each simple object $U\in\text{ob}(\cat)$ we assign a hermitian vector bundle $V_U$ with connection, and to each morphism in $\cat$ we assign a morphism (\ref{eq:2morphisms}) of vector bundles. Moreover, the monoidal structure $J$ describes a family of natural isomorphisms of vector bundles
\begin{equation}\label{eq:gerbeactiongpnatiso}
    J_{X,Y}: V_X\otimes V_Y\xrightarrow[]{\sim} V_{X\otimes Y},
\end{equation}
satisfying the diagram (\ref{eq:relate-assoc}), hence what is understood as a trivialization of the image of the associator in $\cat$, realized as a natural isomorphism of vector bundles, 
\begin{equation}\label{eq:gerbeaction}
    F(a_{X,Y,Z})\circ J_{X\otimes Y,Z}\circ (J_{X,Y}\otimes\text{id}_{U_Z}) = J_{X,Y\otimes Z}\circ (\text{id}_{U_X}\otimes J_{Y,Z}),
\end{equation}
since $\text{HVBdl}_{\nabla}(M)$ is strictly associative. 

A good consistency check is the description of the pure gauge action of finite groups on gerbes. Indeed, the statements above say that a finite group $G$ acts through its comodule category\footnote{Note that since non-isomorphic Hopf algebras may have equivalent representation categories (e.g. the group algebras of isocategorical groups \cite{EG01}), non-isomorphic Hopf algebras may act identically on a given gerbe.}:
\begin{equation}\label{eq:gerbeactiongroup}
    F: \text{Comod}(\C[G])\cong\text{Vec}_G\to \text{HVBdl}_{\nabla}(M).
\end{equation}
Assuming without loss of generality that the line bundle gerbe $\mathcal{L}$ is subordinate to some open cover $\{U_i\}_{i\in \mathcal{I}}$, from Equation (\ref{eq:twistedvector}) we see that the line bundle $L_g$ assigned to each simple $U_g$ is characterized by some Čech 2-cochains labeled by the group elements $g\in G$ 
\begin{equation*}
    1 = \nu_{\alpha\beta}^g\nu_{\beta\gamma}^g\nu_{\gamma\alpha}^g,
\end{equation*}
This agrees with the requirement imposed in \cite[Equation 6]{Sharpe:2000ki} for the special case of trivial action on the base, which translates to $g^*h_{\alpha\beta\gamma}=h_{\alpha\beta\gamma}$ where $h_{\alpha\beta\gamma}$ is the Čech cocycle that determines the bundle gerbe. Moreover, the monoidal structure condition (\ref{eq:gerbeaction}) on the simples implies that the natural isomorphisms (\ref{eq:gerbeactiongpnatiso}) described by cochains (\ref{eq:2morphisms}) are actually Čech 2-cocycles, as deduced in \cite[Equation 7,8]{Sharpe:2000ki} for trivial $g^*$ pullbacks
\begin{eqnarray*}
    \nu_{\alpha\beta}^{g_1g_2} &=& \nu_{\alpha\beta}^{g_2}\,\nu_{\alpha\beta}^{g_1}\,h_{\alpha}^{g_1,g_2} (h_{\beta}^{g_1,g_2})^{-1},\\
    (h^{g_1,g_2g_3}_{\alpha})(h^{g_2,g_3}_{\alpha}) &=& (h^{g_1,g_2}_{\alpha})(h^{g_1g_2,g_3}_{\alpha}).
\end{eqnarray*}
The information about the action on the connective structure as listed in \cite[p.8]{Sharpe:2000ki} is also determined by the monoidal functor, as one can readily verify in a way analogous to what we just presented. However, the tensor functor also determines the action of the non-invertible objects. The functor states that the non-invertible gauge transformation associated with a non-invertible object $\mathcal{O}$ in $\cat$, which by semisimplicity of $\cat$ is isomorphic to a sum of simples
\begin{equation}
    \mathcal{O}\cong \bigoplus_{g\in G}\,N^g_{\mathcal{O}}U_g,
\end{equation}
for $N_{\mathcal{O}}^g\in\mathbb{N}$, is realized by a hermitian vector bundle $V_{\mathcal{O}}$ with connection that is isomorphic to the Whitney sum of the $L_g$ line bundles assigned to the simples
\begin{equation}
    V_{\mathcal{O}}=F(\mathcal{O})\cong \bigoplus_{g\in G}N^g_{\mathcal{O}}L_g,
\end{equation}
something that is expected but not visible by only looking at invertible actions of $G$ on a $U(1)$ bundle gerbe. This shows that (\ref{eq:gerbeactiongroup}) is the correct linear extension of a group action on a bundle gerbe with connection with trivial action on the base.

An immediate generalization of the situation above is the description of actions of (finite-dimensional semisimple) Hopf algebras $\mathcal{H}$, which again acts through its comodule category:
\begin{equation}
    (F,J):\text{Rep}(\mathcal{H}^*)\to \text{HVBdl}_{\nabla}(M).
\end{equation}
Let us briefly comment on the existence of these actions. One can see that in general there are more fusion categories admitting pure gauge actions on a gerbe than fusion categories admitting a fiber functor $(U,J)$ to $\text{Hilb}_{\rm f.d.}$ the category of finite-dimensional Hilbert spaces. This is because if a category admits such a fiber functor $(U,J)$, then it admits an action on any gerbe via the inclusion
\begin{eqnarray}
    \text{Hilb}_{\rm f.d.}\hookrightarrow \text{HVBdl}_{\nabla}(M), \label{eq:hilbinclusion}\\
    H\mapsto M\times H, \label{eq:hilbinclusionob}
\end{eqnarray}
with trivial connective structure. However, an arbitrary action $(F,J):\cat\to\text{HVBdl}_{\nabla}(M)$ does not necessarily factor through $\text{Hilb}_{\rm f.d.}$.

While admitting a pure gauge action on a gerbe is thus a weaker requirement than admitting a fiber functor to $\text{Hilb}_{\rm f.d.}$, many common categories still do not admit actions of this form. This situation can be noticeably improved by also admitting quasi-monoidal functors, which as explained in Section~\ref{ssec:anomalies} amounts to allowing for anomalous actions. Yet, fusion categories with objects with non-integer quantum dimensions still do not admit an action in this sense, such as the Ising category which contains a non-invertible simple object, the Kramers-Wannier duality defect, whose quantum dimension is $\sqrt{2}$. This indicates that some refinements to the notion of action are necessary, including a nontrivial action on the phase space, and the description of fully quantum (as opposed to prequantum) actions.

Finally, we come to the role of discrete torsion twists on these actions. Assuming the fusion category $\cat$ admits a pure gauge action $(F,J)$ on a gerbe, one can now act (\ref{eq:functoraction}) by discrete torsion twists to get a different action. Thus, here the \textit{choice} of structure is the action, on which discrete torsion twists act. Since the twist by discrete torsion does not change the underlying functor $F$, the vector bundles assigned by the twisted functor do not change. On the other hand, the monoidal structure does change, and in particular one obtains morphisms of vector bundles
\begin{equation}
    F(\omega_{X,Y}): V_{X\otimes Y}\xrightarrow{\cong} V_{X\otimes Y},
\end{equation}
for each natural isomorphism $\omega_{X,Y}$ in $\cat$. In this sense, in the context of 2d QFT's coming from $\sigma$-models on line bundle gerbes with connection, the discrete torsion twists \textit{acts} (but \textit{is not}) some given $\cat$-action $(F,J)$ to the action $(F,F\omega\circ J)$, and is observed on the gerbe as a natural collection of invertible 1-form symmetries $F\omega_{X,Y}$ that depends (via whiskering) on a prescribed 0- and 1-form $\cat$-symmetry $(F,\omega)$. This collection further satisfies a 2-cocycle condition described by the image of (\ref{eq:naturalnorm}),(\ref{eq:explicitassoc}) under $F$.

We finish this subsection by specializing to the case when the gerbe is subordinate to a cover, and $\cat=\text{Vec}_G$. In this setting, the morphisms of vector bundles for the simples are the Čech cochains $\omega^{g,h}_{\alpha}$ assigned to each group element described in \cite[Equation 14]{Sharpe:2000ki}:
\begin{eqnarray}\label{eq:gerbeactiondifference}
    \omega^{g,h}_{\alpha} &=& \frac{h^{g,h}_{\alpha}}{\overline{h}^{g,h}_{\alpha}},
\end{eqnarray}
where $h^{g,h}$ and $\overline{h}^{g,h}$ are natural isomorphisms $J_{g,h}$ of the twisted and original action functors, respectively. 

One should note the difference between the derivation in the present article and that in \cite{Sharpe:2000ki}. In the cited article, the cochains (\ref{eq:gerbeactiondifference}) are defined as a ratio of two arbitrary group actions (possibly acting nontrivially on the base), and it is then observed that these cochains are actually cocycles. Here, by contrast, that such cochains are cocycles follows automatically, since we are considering the ratio of some (pure gauge) group action $F$ and its twist by a 2-cocycle $\omega$ in $\text{Vec}_G$, which by construction is the whiskering $F(\omega)$. Note moreover that if the action $(F,J)$ of $\text{Vec}_G$ factors through $\text{Hilb}_{\rm f.d.}$ (\ref{eq:hilbinclusion},~\ref{eq:hilbinclusionob}), the transition functions in $\text{HVBdl}_{\nabla}(M)$ assigned to the components $F(\omega_{g,h})$ of the simples are genuine group 2-cocycles classified in $H^2(G,U(1))$, which is the specialization to topologically-trivial principal $U(1)$ bundles described in \cite[Section 3.3]{Sharpe:2000ki}.

\subsection{Anomalies and quasi-monoidal functors}\label{ssec:anomalies}

We finalize this section by exploring some of the consequences of relaxing the notion of a monoidal functor to that of a quasi-monoidal functor by dropping the condition (\ref{eq:relate-assoc}). This means that a quasi-monoidal functor is again a pair
\begin{equation}\label{eq:quasifiber}
    (Q,j):\cat\to\mathcal{E},
\end{equation}
where $Q$ is a functor of underlying categories, and $j$ is a natural isomorphism (\ref{eq:jnaturaliso})
\begin{equation*}
   \begin{tikzcd}
\mathcal{C}\times \mathcal{C} \arrow[rrr, "\otimes_{\mathcal{C}}"] \arrow[dd, "Q\times Q"']        &  &  & \mathcal{C} \arrow[dd, "Q"] \\
                                                                                                   &  &  &                             \\
\mathcal{E}\times \mathcal{E} \arrow[rrr, "\otimes_{\mathcal{E}}"'] \arrow[rrruu, "j", Rightarrow] &  &  & \mathcal{E}                
\end{tikzcd}
\end{equation*}
without further requirements. These functors, as we describe below, are intimately related to the notion of anomalies. In particular, we will see that one can have discrete torsion twists without a fiber functor.

The important observation for discrete torsion is that given a quasi-monoidal functor $(Q,j)$ one may also twist it by a 2-cocycle $\omega$ to obtain another quasi-monoidal functor $(Q,Q(\omega)\circ j)$, which as one can readily verify is monoidal if and only if $(Q,j)$ itself is monoidal. This allows to extend the notion of discrete torsion twists and actions to categories of representations of quasi-Hopf algebras, for instance. We now illustrate this concretely.

A first example is provided by considering quasi-monoidal functors in the context of Section~\ref{ssec:fiberfunc}. For that purpose, consider the category $\text{Vec}_G^{\alpha}$ of $G$-graded vector spaces whose associator is now defined by a nontrivial cohomology class $0\neq [\omega]\in H^3(G,\C^{\times})$. It is well known that this fusion category does not admit a fiber functor, equivalently an action on $\text{Vec}$, and for this reason one calls the category $\text{Vec}_G^{\alpha}$ anomalous \cite[Section 2.3]{Thorngren:2019iar}. However, it is still a fusion category and as such it has a well-defined lazy cohomology group, since it is an intrinsic notion. Indeed, its discrete torsion twist group $H^2_{\ell}(\text{Vec}_G^{\alpha})$ is the same as for the category $\text{Vec}_G$, that is, $H^2(G,U(1))$. These lazy 2-cocycles, in the same fashion described in (\ref{eq:moritaaction}), can then twist any gaugeable algebra in $\text{Vec}_G^{\alpha}$ such as the algebras corresponding to the subgroups $H\leq G$ for which $\alpha\vert_H$ is cohomologically trivial.

Now, a lazy 2-cocycle can no longer give a fiber functor, since there are no fiber functors on which it can act. However, $\text{Vec}_G^{\alpha}$ does admit a quasi-fiber functor \cite[Example 5.1.3]{EGNO}. This can consist, for example, of the forgetful functor of underlying categories that ignores the $G$-grading, along with the identity natural isomorphism
\begin{gather}
    Q: \text{Vec}_G^{\alpha}\to \text{Vec},
    \\
    j_{X,Y}: Q(X)\otimes Q(Y)\xrightarrow{\text{id}} Q(X\otimes Y).
\end{gather}
The anomaly is then observed in $\text{Vec}$ as the non-associativity of the images of the objects of $\text{Vec}_G^{\alpha}$ in $\text{Vec}$, equivalently their actions on $\text{Vec}$. More in detail, $(Q,j)$ defines a diagram for all triples of objects
\begin{equation}   \label{eq:quasirelate-assoc}
\begin{tikzcd}
(Q(X)\otimes Q(Y))\otimes Q(Z) \arrow[rrr, "{\Phi_{X,Y,Z}}"] \arrow[d, "{j_{X,Y}\otimes \text{id}_{Q(Z)}}"'] &  &  & Q(X)\otimes (Q(Y)\otimes Q(Z)) \arrow[d, "{\text{id}_{Q(X)}\otimes j_{Y,Z}}"] \\
Q(X\otimes Y)\otimes Q(Z) \arrow[d, "{j_{X\otimes Y,Z}}"']                                                         &  &  & Q(X)\otimes Q(Y\otimes Z) \arrow[d, "{j_{X,Y\otimes Z}}"]                     \\
Q((X\otimes Y)\otimes Z) \arrow[rrr, "{Q(a_{X,Y,Z})}"']                                                            &  &  & Q(X\otimes (Y\otimes Z))                                                     
\end{tikzcd}.
\end{equation}
where $\Phi_{X,Y,Z}$ is generally a nontrivial morphism. Concentrating on triples of simple objects $(U_g,U_h,U_z)$ gives morphisms $\Phi_{g,h,k}:\C\to\C$, which is simply the value of the 3-cocycle associator $\alpha(g,h,k)$. The reader should compare this to the case of fiber functors (\ref{eq:relate-assoc}) to $\text{Vec}$, in which case $\Phi_{X,Y,Z}=a_{F(X),F(Y),F(Z)}=\text{id}_{F(X)\otimes F(Y)\otimes F(Z)}$.

Having a quasi-fiber functor at our disposal, one can proceed and twist by a lazy 2-cocycle, which returns another quasi-fiber functor. This may be understood as discrete torsion twists of anomalous actions on $\text{Vec}$. 

We can also consider, as in Section \ref{ssec:fiberfunc}, relaxing the condition on the action functor from a monoidal functor to a quasi-monoidal functor $(Q,j)$ for actions of gerbes, which we understand as anomalous actions. As in the case above, an anomalous action gives rise to diagrams
\begin{equation} 
\begin{tikzcd}
(Q(X)\otimes Q(Y))\otimes Q(Z) \arrow[rrr, "{\Phi_{X,Y,Z}}"] \arrow[d, "{j_{X,Y}\otimes \text{id}_{Q(Z)}}"'] &  &  & Q(X)\otimes (Q(Y)\otimes Q(Z)) \arrow[d, "{\text{id}_{Q(X)}\otimes j_{Y,Z}}"] \\
Q(X\otimes Y)\otimes Q(Z) \arrow[d, "{j_{X\otimes Y,Z}}"']                                                         &  &  & Q(X)\otimes Q(Y\otimes Z) \arrow[d, "{j_{X,Y\otimes Z}}"]                     \\
Q((X\otimes Y)\otimes Z) \arrow[rrr, "{Q(a_{X,Y,Z})}"']                                                            &  &  & Q(X\otimes (Y\otimes Z))                                                     
\end{tikzcd},
\end{equation}
where the collection of such $\Phi_{X,Y,Z}$ are now potentially non-trivial morphisms of hermitian vector bundles with connection
\begin{equation}\label{eq:categoryanomaly}
    \Phi_{X,Y,Z}: Q(X)\otimes Q(Y)\otimes Q(Z) \xrightarrow[]{\cong} Q(X)\otimes Q(Y)\otimes Q(Z),
\end{equation}
where we omit the brackets since $\text{HVBdl}_{\nabla}(M)$ is strictly associative. The collection of isomorphisms (\ref{eq:categoryanomaly}) characterizes the anomaly of the $\cat$-action.

In particular, for $\cat=\text{Vec}_G^{\alpha}$ for $0\neq[\alpha]\in H^3(G,U(1))$, we can see that an anomalous action on a gerbe is described, apart from the cochains coming from the natural isomorphisms $j_{X,Y}$, by Čech 3-cocycles
\begin{equation}
    (k^{g_2,g_3,g_4}_{\alpha})(k_{\alpha}^{g_1,g_2g_3,g_4})(k_{\alpha}^{g_1,g_2,g_3}) = (k_{\alpha}^{g_1g_2,g_3,g_4})(k_{\alpha}^{g_1,g_2,g_3g_4}).
\end{equation}
If it furthermore happens that this action factors (\ref{eq:hilbinclusion},~\ref{eq:hilbinclusionob}) through the trivial finite-dimensional Hilbert bundles $M\times H$, hence factors through a quasi-fiber functor to $\text{Hilb}_{\rm f.d.}$, then these are genuine group $3$-cocycles as the ones classifying the 't Hooft anomaly of some $G$-symmetry.

Just as before, these anomalous actions described by quasi-monoidal functors can be acted on by the intrinsic discrete torsion twists of $\cat$ to yield other anomalous actions. This shows that the intrinsic formulation of discrete torsion twists seamlessly allows for significant generalizations of discrete torsion action.

\section{Conclusions}

In this article we have highlighted that the notion of discrete notion of group-like symmetries actually admits two different but complementary generalizations to non-invertible symmetries: the \textit{set} of discrete torsion \textit{choices}, and the \textit{cohomology group} of discrete torsion \textit{twists}. These two notions are complementary in the sense that the former always admits a consistent \textit{twisting action} by the latter, meaning that discrete torsion choices not only form a set but a group-module. We showed that the discrete torsion twists of a fusion category $\cat$ are coherently described by its lazy cohomology group $H^2_{\ell}(\cat)$. We explicitly described how lazy 2-cocycles can consistently twist the gaugeable algebras $\mathcal{A}$ inside $\cat$ to produce new gaugeable algebras, and that up to Morita equivalence the twist only depends on the cohomology class of the lazy 2-cocycle. We gave explicit expressions for the resulting partition functions, and showed they are consistent with modular invariance and multi-loop factorization. We also described how lazy 2-cocycles can twist monoidal functors from $\cat$, where similarly only lazy 2-cocycles with nontrivial cohomology classes can give rise to monoidal inequivalent twists. In particular, we studied the case of twists of fiber functors to clarify the distinction between discrete torsion twists, and discrete torsion choices, described by equivalence classes of fiber functors. We also studied how discrete torsion twists relate to non-invertible gauge actions on B fields, and recovered known results in the special case of group-like actions. The perspective of separating between choices of structure and twists of such structures thus allows a direct application of discrete torsion twisting actions to a wide variety of settings, including actions on $\text{Vec}$ (fiber functors), but also explicit actions on gerbes and even their anomalous variants.

\section{Acknowledgements}
We would like to thank D. Robbins, E. Sharpe, T. Vandermeulen, and X. Yu for useful discussions.

\appendix

\section{Lazy cohomology}\label{app:lazycohomology}
For the reader's convenience, we gather the definitions relevant to the characterization of discrete torsion twists as elements of the lazy cohomology group.

Let $\cat$ be a fusion category with its tensor product functor denoted as $\otimes$. A \textit{lazy 2-cocycle} is a natural isomorphism
\begin{equation*}
    \omega: \otimes \Rightarrow \otimes,
\end{equation*}
satisfying the conditions
\begin{gather*}
    \omega_{X,1}=\omega_{1,X} = \text{id}_X;
    \\
    \omega_{X,Y\otimes Z} \circ (\text{id}_X\otimes \omega_{Y,Z}) = \omega_{X\otimes Y,Z}\circ (\omega_{X,Y}\otimes \text{id}_Z).
\end{gather*}
The collection of lazy 2-cocycles form a group denoted as $Z^2(\cat)$.

A \textit{lazy 2-coboundary} is a lazy 2-cocycle whose components are of the form
\begin{equation*}
    (d\phi)_{X,Y}:=\phi^{-1}_{X\otimes Y}\circ (\phi_X\otimes\phi_Y),
\end{equation*}
where $\phi$ is an automorphism of $\cat$
\begin{equation}
    \phi: \text{id}_{\mathcal{C}}\to \text{id}_{\mathcal{C}}.
\end{equation}

The \textit{lazy cohomology} $H^2_{\ell}(\cat)$ of $\cat$ is the quotient of lazy 2-cocycles $Z^2(\cat)$ by its central subgroup $B^2(\cat)$ of lazy 2-coboundaries 
\begin{equation*}
    H^2_{\ell}(\mathcal{C}) := Z^2(\mathcal{C})/B^2(\mathcal{C}).
\end{equation*}

The cohomology classes in $H^2_{\ell}(\cat)$ is what we identify as discrete torsion twists.

\end{document}